\documentclass[aps,pra,twocolumn,amsmath,amssymb,superscriptaddress,reprint,longbibliography]{revtex4-2}

\usepackage[utf8]{inputenc}
\usepackage{graphicx}
\usepackage[colorlinks=true,citecolor=blue]{hyperref}
\usepackage[caption=false]{subfig}
\usepackage{lipsum}

\begin{document}

\title{Modeling language ideologies for the dynamics of languages in contact}
\author{Pablo Rosillo-Rodes}
\email{prosillo@ifisc.uib-csic.es}
\author{Maxi San Miguel}
\author{David Sánchez}

\affiliation{
 Institute for Cross-Disciplinary Physics and Complex Systems IFISC (UIB-CSIC), Campus Universitat de les Illes Balears, E-07122 Palma de Mallorca, Spain
}
\date{\today}

\begin{abstract}
In multilingual societies, it is common to encounter different language varieties. Various approaches have been proposed to discuss different mechanisms of language shift. However, current models exploring language shift in languages in contact often overlook the influence of language ideologies. Language ideologies play a crucial role in understanding language usage within a cultural community, encompassing shared beliefs, assumptions, and feelings towards specific language forms. These ideologies shed light on the social perceptions of different language varieties expressed as language attitudes. In this study, we introduce an approach that incorporates language ideologies into a model for contact varieties by considering speaker preferences as a parameter. Our findings highlight the significance of preference in language shift, which can even outweigh the influence of language prestige associated, for example, with a standard variety. Furthermore, we investigate the impact of the degree of interaction between individuals holding opposing preferences on the language shift process. Quite expectedly, our results indicate that when communities with different preferences mix, the coexistence of language varieties becomes less likely. However, variations in the degree of interaction between individuals with contrary preferences notably lead to non-trivial transitions from states of coexistence of varieties to the extinction of a given variety, followed by a return to coexistence, ultimately culminating in the dominance of the previously extinct variety. By studying finite-size effects, we observe that the duration of coexistence states increases exponentially with network size. Ultimately, our work constitutes a quantitative approach to the study of language ideologies in sociolinguistics.
\end{abstract}

\maketitle

\textbf{Languages come in different forms or varieties, making them diverse and interesting. The way people speak a language depends on various factors like how highly it is regarded in society, which can determine whether it survives or disappears over time. Additionally, individual speakers often have their own preferences for specific language varieties, which can balance out the influence of societal views. To better understand how the use of these language varieties evolves, we develop a dynamic model that considers both personal preferences and societal opinions. Our research shows that when communities with opposing language preferences are more interconnected, it becomes challenging for different varieties to coexist. These findings could have important implications for policies aimed at preserving endangered languages.}

\section{Introduction}
Modeling language shift is valuable because it can unveil the mechanisms
that lead to language death or its maintenance~\cite{Krauss1992,CrystalDeath,Mufwene2004}.
The pioneering model of Abrams
and Strogatz~\cite{Abrams2003} assumed that language shift is mostly driven by a prestige parameter,
which quantifies the relative strength between two linguistic varieties
in contact but with different sociolinguistic statuses~\cite{Chambers1998}.
When the transition rate for speakers to change their initial language
is proportional to the number of people that speak the target language, the only stable fixed point of the model implies
the extinction of the variety with the least prestige.
Interestingly, the extinction processes of languages have analogs with the evolutionary
properties of biological species~\cite{Lieberman2007,Atkinson2008,Steele2010}.

Since then, different
mechanisms~\cite{Patriarca2004,Mira2005,Castello2006,Minett2008,Patriarca2009,Kandler2010,Patriarca2012,Isern2014,Prochazka2017,Luck2020,Uriarte2021} have been proposed to enable the coexistence of varieties seen in reality, 
which in fact is a rather common situation in multilingual societies~\cite{Maffi2005,Fincher2008,Louf2021,Seoane2022}.
For instance, a community of bilingual speakers may help stabilize a fixed
point with different fractions of monolingual speakers. Another possibility
is to introduce a volatility parameter, which accounts for the fact
that a speech community can be more opaque to the influence of speakers
with a different variety. However, all these theoretical approaches
(reviewed in, e.g., Refs.~\cite{Wang2005,Sole2010,Baronchelli2012,Boissonneault2021})
do not fully take into account the role of language ideologies,
a social factor that is currently considered as
as a key concept in understanding
language use and attitudes within a cultural group.
This is the gap we want to fill in with our work.

Language ideologies comprise a wide spectrum of beliefs, assumptions
and feelings that a group of speakers socially share about certain
language forms~\cite{AlburyAttitudes}.
As such, ideologies lead to linguistic attitudes~\cite{Garrett2001,GarrettAttitudesBook} and values that express
with explicit actions degrees of favor or disfavor toward a language or a variety.
These psychological tendencies generate prejudices, stereotypes, biases, etc.
A commonplace case refers to languages that have undergone a standardization
process in which the standard variety is advocated in school, government offices
and mass media against the vernacular variety or dialect spoken in a particular
region~\cite{MilroyIdeologyStandard}. Typically, this leads to an overt prestige that encourages speakers
to use the standard variety by penalizing utterances that
depart from the linguistic norms. However, there also exists a covert prestige~\cite{LabovSP}
that describes a positive willingness towards socially considered lower forms
due to cultural attachment or group identity with regard to the vernacular
variety. This can happen owing to the presence of ethnic differences
(e.g., African-American English~\cite{White1998}) or the influence of a third variety
(e.g., bilingual Basque-Spanish speakers preferring on average Basque Spanish to Standard Spanish~\cite{Elordieta2021}),
among other causes. From the viewpoint of mathematical modeling, an equivalent
situation considers the competition between a global and a local language
(the latter may be endangered), where these two languages are related vis-à-vis
with the standard and vernacular varieties indicated above. Further, one could
envisage two ways of speaking (young versus old generations, high versus
low socioeconomic classes, etc.) associated to distinct sociological parameters.
Our theoretical proposal is thus completely general in this respect and just
considers two speech communities with different linguistic preferences
and two language varieties in contact with different prestige. This way our findings
can be applied to a broad range of sociolinguistic situations.

Our model builds upon previous efforts~\cite{Liggett1985,Castellano2009,Masuda2010,Masuda2011,Baronchelli2018,Redner2019} that 
consider communities of binary agents with different states. The agents can change
their states interacting with their neighbors following predefined rules.
As a consequence, the state of the population evolves in time until a consensus
is reached (or not). In our case, the state is the language or variety
spoken by the agent while the transition rates for variety adoption reflect
the influence of the surrounding individuals in terms of the variety prestige
and the fraction of those individuals speaking any of the two varieties.
Crucially, the agents can have two internal preferences caused by their language ideologies, distinct from their state concerning the spoken variety. Consequently, agents may prefer either their spoken variety or the alternative existing one. These preferences for the standard or the
vernacular variety determine in term the
values assigned to each variety prestige. In short,
the model accounts not only for what language the individuals speak
but also what language they prefer to speak. Our findings reveal that in some cases
the agents' preference can counteract the force of the most prestigious variety,
thus leading to the survival (or even dominance) of the local variety in relation
with the standard variety. More strikingly, our model
shows a rich constellation of phases---upon increasing of the coupling
between the two communities with different preferences we find a transition
from a social state where the vernacular (majority) language dominates to a
phase where this variety becomes extinct,
sandwiched between intermediate regions for which the coexistence between varieties is possible, and finally a phase there the standard (minority) language is dominant
across the society. These results can be better understood in the mean field limit
where the agents are connected all to all. Yet we also investigate finite
size effects with the aid of agent-based modeling and calculate the survival times.
Below, we give more details on this complex landscape,
which both deepens our knowledge on the dynamics of languages in contact and 
and may have an impact in the design of appropriate language policies
that seek to revitalize endangered languages.

\section{Model}
\label{sec:firstmodel}

\begin{figure}[t]
\centering
\includegraphics[width=0.485\textwidth]{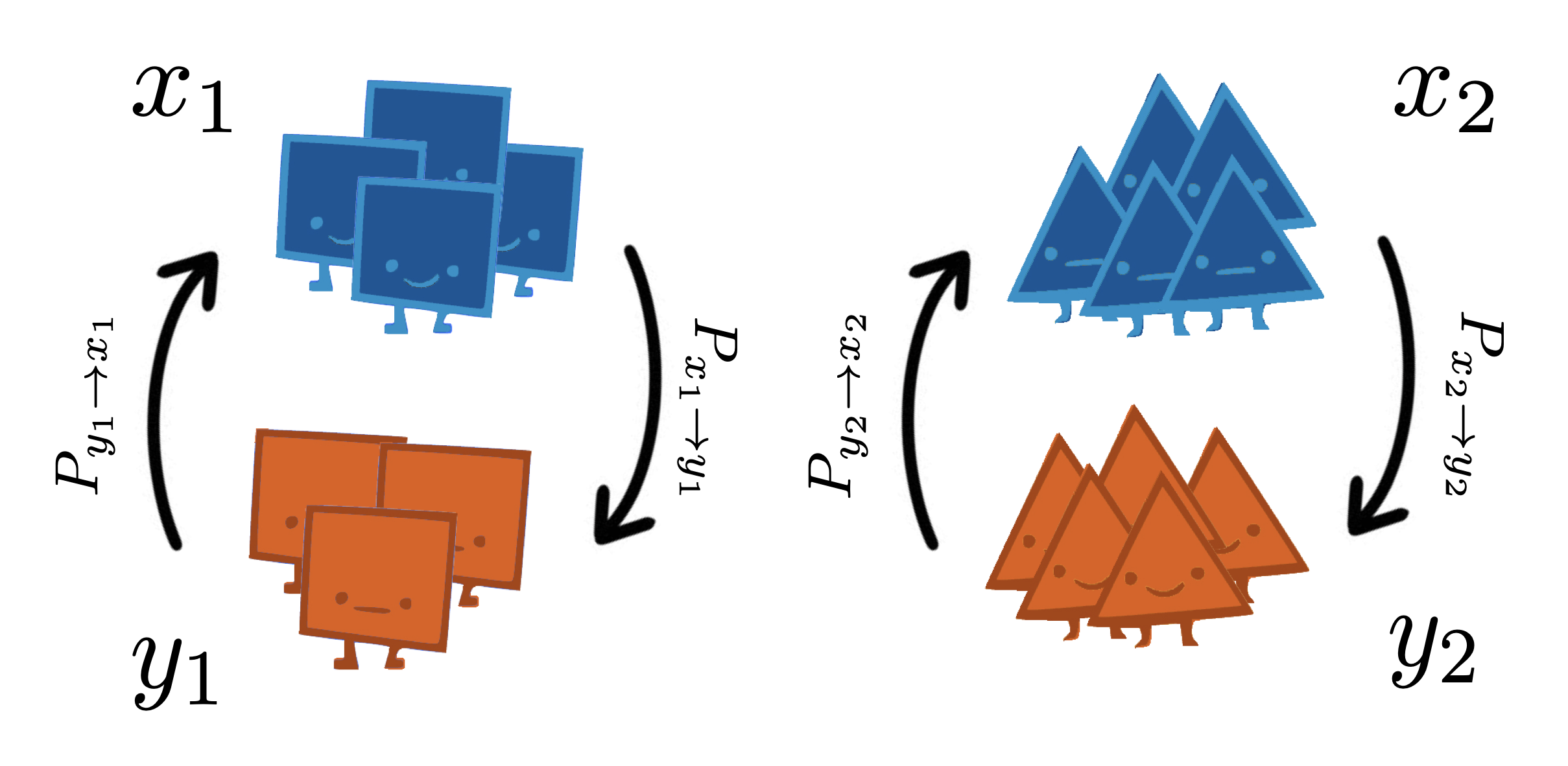}
\caption{Schematic illustration of the model with fixed preferences. The society consists of a large number of speakers who engage in communication with all the other members. Speakers have a fixed internal preference (denoted by their shape, square or triangle) and may speak any of the two varieties (denoted by their color, orange or blue). As their preference is fixed but their spoken variety may change, transitions only occur between population groups with the same preference, i.e., $x_1 \leftrightarrow y_1$ and $x_2 \leftrightarrow y_2$. These population groups may have different sizes, determined by the preference parameter $\alpha = x_1+y_1 = 1-x_2-y_2$. The polygon idea is inspired by Ref.~\cite{polygonparable}.}
\label{fig:scheme}
\end{figure}

Our goal is to quantify the influence that the linguistic preference of the speakers may have over the distribution of speakers within the different varieties of a language. For this purpose, we propose a mathematical framework which models a society in which only one language with two different varieties---the standard and the vernacular---exists.
As explained in the Introduction, this model is also valid for two languages or for two ways of speaking induced by sociological factors.
Therefore, speakers may speak either one variety or the other but they may prefer one variety over the other. 
Let $X$ ($Y$) denote the standard (vernacular) variety while the preference is labeled with $1$ or $2$. This implies that we have four groups of speakers: $x_1$, $x_2$, $y_1$, and $y_2$.
On the one hand, $x_1$ is the fraction of standard speakers that prefer the standard variety whereas $x_2$ is the fraction of standard speakers that prefer the vernacular variety. On the other hand, $y_2$ is the fraction of vernacular speakers that prefer the vernacular variety whereas $y_1$ is is the fraction of vernacular speakers that prefer the standard variety.
This depiction of a society with one language, two varieties and four population groups is the minimal model that captures essentially the influence that preferences have on language shift.

Since we are dealing with population fractions, it is a good approach to consider that our society consists of a large number of interacting speakers. The dynamics of the system when the speakers interact all to all (mean field approximation) is given by the rate equations
\begin{align}
    \frac{\mathrm{d}x_1}{\mathrm{d}t} &= P_{y_1 \rightarrow x_1} y_1 - P_{x_1 \rightarrow y_1} x_1,\label{eq:s1}\\
    \frac{\mathrm{d}x_2}{\mathrm{d}t} &= P_{y_2 \rightarrow x_2} y_2 - P_{x_2 \rightarrow y_2} x_2\label{eq:s2},\\
    \frac{\mathrm{d}y_1}{\mathrm{d}t} &= P_{x_1 \rightarrow y_1} x_1 - P_{y_1 \rightarrow x_1} y_1,\label{eq:s3}\\
    \frac{\mathrm{d}y_2}{\mathrm{d}t} &= P_{x_2 \rightarrow y_2} x_2 - P_{y_2 \rightarrow x_2} y_2,\label{eq:s4}
\end{align}
where the transition rates to shift from variety $X$ ($Y$) to variety $Y$ ($X$) are accordingly proportional to the total number of $Y$ ($X$) speakers:
\begin{align}
    \label{eq:tp1}
    P_{x_1 \rightarrow y_1} &= (1-s_1)(y_1+y_2),\\
    \label{eq:tp2}
    P_{x_2 \rightarrow y_2} &= s_2(y_1+y_2),\\
    \label{eq:tp3}
    P_{y_1 \rightarrow x_1} &= s_1(x_1+x_2),\\
    \label{eq:tp4}
    P_{y_2 \rightarrow x_2} &= (1-s_2)(x_1+x_2).
\end{align}
Importantly, the shift probabilities given by Eqs.~\eqref{eq:tp1}, \eqref{eq:tp2}, \eqref{eq:tp3}, and \eqref{eq:tp4}
include the parameters $s_1$ and $s_2$, which account for the prestige of the standard variety for the vernacular speakers and vice versa. Quite generally, we take
$s_1,s_2>0.5$ to model the fact that those speakers whose preference is not aligned with
their language switch more easily than those speakers whose preference is aligned.
For instance, vernacular speakers who prefer to speak language $X$ (i.e., the group $y_1$) change with a rate proportional to $s_1$ [Eq.~\eqref{eq:tp3}] whereas standard speakers
who speak their preferred language (i.e, the group $x_1$) change with a smaller rate, since this is proportional to $1-s_1$ [Eq.~\eqref{eq:tp1}]. This ingredient is absent from previous
models and emphasizes the importance of preference alignment or disalignment in language shift processes.

On the other hand, 
we take $s_1>s_2$ to reflect the fact that overt prestige, associated to the higher-status language or standarized variety, is higher than covert prestige, associated to the lower-status language or vernacular variety.
However, the mechanism for preference alignment operates similarly as before: those speakers who prefer variety $Y$ (i.e., the group $x_2$) are more likely to shift [Eq.~\eqref{eq:tp2}] that those vernacular speakers whose preference agrees with their variety (i.e., the group $y_2$), see Eq.~\eqref{eq:tp4}.

The fractions in Eqs.~\eqref{eq:s1}, \eqref{eq:s2}, \eqref{eq:s3}, and \eqref{eq:s4} obviously obey
\begin{equation}
    \label{eq:ligadura1}
    x_1 + x_2 + y_1 + y_2 = 1.
\end{equation}
We note that transitions are not allowed
between groups of different preferences. Thus, Eqs.~\eqref{eq:s1}, \eqref{eq:s2}, \eqref{eq:s3}, and \eqref{eq:s4} constitute a {\em fixed-preference} model, meaning a fixed proportion of speakers with a preference for each of the varieties. In Fig.~\ref{fig:scheme} we illustrate the transitions between the different population groups $x_1, y_1, x_2$ and $y_2$, which occur only between groups of speakers with the same preference, i.e., $x_1 \leftrightarrow y_1$ and $x_2 \leftrightarrow y_2$.

This is especially relevant for populations that may change their language but not their preference. Of course, preferences can evolve with time but language ideologies
are maintained in a population typically over a generation~\cite{McIntosh2014}, much longer
than language change of usage, which can occur at a significantly higher rate~\cite{So2013}. 
Therefore, our results are restricted to time ranges when language shift can take place
but preferences are constant.

We define the constant $\alpha$
as the total fraction of speakers who prefer the standard variety, 
\begin{equation}
    \label{eq:ligadura2}
    \alpha = x_1 + y_1,
\end{equation}
Using Eq.~\eqref{eq:ligadura1} this also determines the fraction of speakers
who prefer the vernacular variety, $x_2 + y_2=1-\alpha$.

\section{Fixed points}

\begin{table}
\centering
\begin{tabular}{|c|c|c|} \hline
ID & $X^*$ & $\omega^*$   \\ \hline
E  & 0 & 0          \\
D  & 1 & $2\alpha-1$  \\
C &
  $X^*_\mathrm{C}$ [Eq. (\ref{eq:xfp3c})] & $\omega^*_\mathrm{C}$ [Eq. (\ref{eq:wfp3c})]  \\ 

  \hline \hline
ID & $x_1^*$ & $y_2^*$   \\ \hline
E & 0  & $1-\alpha$ \\
D & $\alpha$ & 0 \\
C &
  $\left(X^*_\mathrm{C} + \omega^*_\mathrm{C} \right)/2$ & $\left[2(1-\alpha)-X^*_\mathrm{C}+\omega^*_\mathrm{C}\right]/2$ \\ 

  \hline

\end{tabular}
\caption{Fixed points for the dynamical model with fixed preference $\alpha$. From $X^*$ and $\omega^*$ one can find the value of $Y^* = 1-X^*$ and $z^* = \omega^*-2\alpha+1$. Each fixed point is labelled with an identifier (ID) which refers to the extinction (E) of the standard variety, its dominance (D), or coexistence between standard and vernacular varieties (C).}
\label{tab:fixedpoints}
\end{table}

To understand more easily the results, it is convenient to make the change of variables 
\begin{equation}
    \left(X, Y, \omega, z\right) = \left(x_1+x_2, y_1+y_2, x_1-x_2, y_2-y_1\right),
\end{equation}
where $X$ and $Y$ are clearly the total speakers for standard and vernacular varieties, respectively, and $\omega$ and $z$ quantify how many
speakers of $X$ and $Y$, respectively, are aligned (or antialigned) with their internal preferences. Due to constraints imposed by Eqs.~\eqref{eq:ligadura1} and~\eqref{eq:ligadura2}, our original set of 4 independent variables $x_1$, $x_2$, $y_1$ and $y_1$ turns into a set with only two independent variables, chosen to be $X$ and $\omega$.

Thus, the dynamics of the system are governed by the rate equations
\begin{align}
    \label{eq:xeq}
    \notag
    \frac{\mathrm{d}X}{\mathrm{d}t} &= (s_1-s_2)X(1-X)  \\
    &  -\frac{1}{2}(s_1+s_2-1)\left[(1-2\alpha)X+(2X-1)\omega\right], \\
    \label{eq:weq}
    \notag
    \frac{\mathrm{d}\omega}{\mathrm{d}t} &= \frac{1}{2} \left[ (s_1-s_2-1)\omega+2(s_2-s_1)X\omega \right. \\
    \notag
    &\left.   +2(s_1+s_2-1)X(1-X) \right. \\
    &\left. +(2\alpha-1)(1+s_1-s_2)X \right],
\end{align}
which are the result of combining Eqs. \eqref{eq:s1}, \eqref{eq:s2}, \eqref{eq:ligadura1} and $\eqref{eq:ligadura2}$ properly.

The two terms constituting Eq.~\eqref{eq:xeq} have a clear interpretation. The first corresponds to the logistic equation, with only two fixed points at $X = 0$ and $X = 1$. However, the inclusion of preferences in the second term introduces a new fixed point allowing for coexistence. Due to $s_1 > s_2$, the sign of the first term is always positive. Hence, the sign of $\mathrm{d}X/\mathrm{d}t$ will be given by the second term, since $\left[(1-2\alpha)X+(2X-1)\omega\right]$ may be either positive, negative or null, and $(s_1+s_2-1) < 1$ holds true at all times. This inequality is guaranteed because $s_1, s_2 > 0.5$ and $s_1 > s_2$. 

We show in Table~\ref{tab:fixedpoints} the analytical expressions for the fixed points of Eqs.~\eqref{eq:xeq} and~\eqref{eq:weq} for these two independent variables $X$ and $\omega$.

Fixed points with IDs E and D imply the extinction of one of the varieties: E describes a situation in which all speakers employ the vernacular variety while D implies that all individuals speak the standard variety.
In turn, C implies the coexistence for speakers of both varieties. This is the first remarkable result as compared with Ref.~\cite{Abrams2003}, where coexistence is not possible for linear transition rates. The fraction of speakers of each variety and their preference distribution depends on $s_1$, $s_2$ and $\alpha$. In contrast, extinction and dominance fixed points E and D, respectively, are independent of the parameters of the model. Indeed, they constitute absorbing states in a stochastic simulation. We will later elaborate on this observation when we discuss our agent-based model simulations.

Fixed points of extinction (E) and dominance (D) of the standard variety always exist within the limits of the phase plane, i.e., $0\leq X\leq1$, $-1\leq \omega\leq 1$. However, the fixed point implying coexistence of varieties (C) only lies inside the existence range of the chosen variables when
\begin{equation}
    \label{eq:alpharange}
    \alpha_1 \leq \alpha \leq \alpha_2,
\end{equation}
with
\begin{equation}
    \notag
    \alpha_1 = \frac{(1-s_1)(2s_2-1)}{s_1+s_2-1} \ \mathrm{and} \ \alpha_2 = \frac{s_1 (2s_2-1)}{s_1+s_2-1}.
\end{equation}

As for the stability of the fixed points, the two eigenvalues $\lambda_1$ and $\lambda_2$ of the Jacobian matrix that result from the linearization of the dynamical equations around one of these fixed points are given by Eq.~\eqref{eq:eig1}. A computational analysis of the expressions for the fixed points in Table~\ref{tab:fixedpoints} and their stability following Eq.~\eqref{eq:eig1} yields two important results. First, the eigenvalues $\lambda$ are always real. We can then exclude dynamic states such as cycles. Second, there is always one and only one stable fixed point for each parameter configuration. Depending on the parameters, there will be only one steady state characterised by the extinction of the standard variety (E), its dominance (D), or the coexistence of the two varieties (C). When $\alpha < \alpha_1$, extinction of the standard variety (E) is stable, and when $\alpha > \alpha_2$ standard dominance (D) is stable.

Remarkably, the set of parameters that implies the stability of the coexistence fixed point (C), computed by imposing $\lambda_1 < 0$ and $\lambda_2 < 0$ for $X^* = X^*_\mathrm{C}$ and $\omega^* = \omega^*_\mathrm{C}$, is also given by Eq.~\eqref{eq:alpharange}, meaning that, whenever coexistence is possible, it is stable over time.

\begin{figure*}

\begin{minipage}{.5\linewidth}
\centering
\subfloat{\label{fig:phasespacefp:a}\includegraphics[scale=0.8]{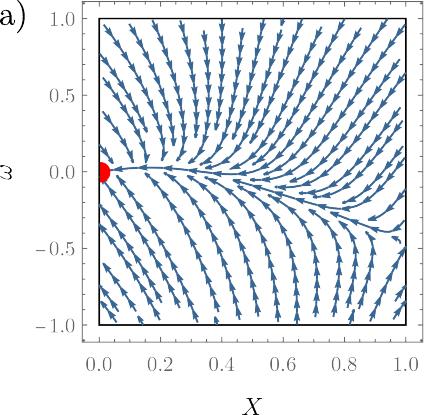}}
\end{minipage}%
\begin{minipage}{.5\linewidth}
\centering
\subfloat{\label{fig:phasespacefp:b}\includegraphics[scale=0.8]{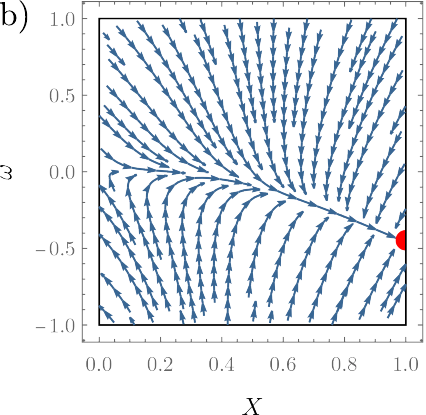}}
\end{minipage}\par\medskip
\begin{minipage}{.5\linewidth}
\centering
\subfloat{\label{fig:phasespacefp:c}\includegraphics[scale=0.8]{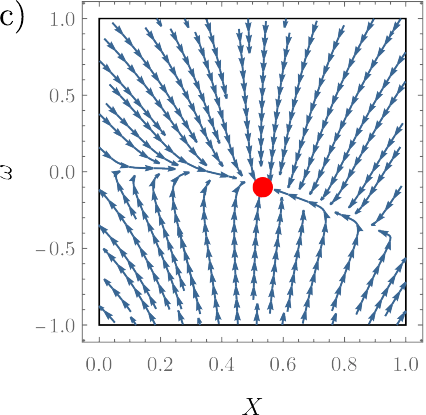}}
\end{minipage}%
\begin{minipage}{.5\linewidth}
\centering
\subfloat{\label{fig:phasespacefp:d}\includegraphics[scale=0.8]{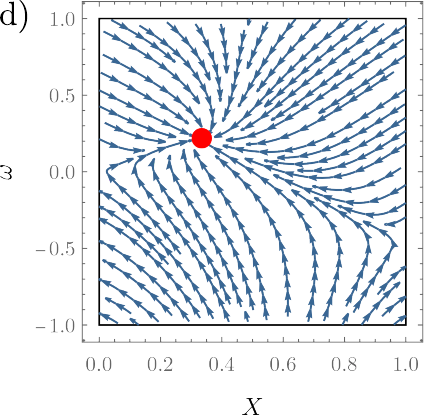}}
\end{minipage}\par\medskip

\caption{Phase portrait for different configurations of the model with preference  $\alpha = 0.27$. In red we plot the location of the stable fixed point. a) For $s_1 = 0.7$, $s_2 = 0.68$ we show a situation of the death for the standard variety since the stable fixed point is located at $X^* = 0$. b) For $s_1 = 0.8$, $s_2 = 0.55$, conversely, we depict the death for the vernacular variety because the stable fixed point implies $X^* = 1$. Both situations c) with $s_1 = 0.8$, $s_2 = 0.6$ and d) with $s_1 = 1$, $s_2 = 0.85$ show coexistence of both varieties.}
\label{fig:phasespacefp}
\end{figure*}

We will now investigate the influence of speakers' preferences in the particular state achieved by the system in the long time limit. As we mentioned before, there is always one and only one stable fixed point for every possible value of the triad $(s_1, s_2, \alpha)$. Figure~\ref{fig:phasespacefp}a) shows an interesting case: despite the fact that the standard variety has a higher prestige, the stable solution corresponds to all speakers using the vernacular variety (E). This is because, for particular values of $\alpha<0.5$, the community preference is biased for the vernacular variety. Consequently, a sufficiently low value of $\alpha$ can counteract the strength of a higher prestige variety. Both $X$ and $\omega$ are null in this case, as all the standard variety speakers switch to the vernacular variety.
In Fig.~\ref{fig:phasespacefp}b) we depict an expected case: if $s_1$ is sufficiently large as compared with $s_2$, the preference $\alpha$ cannot prevent the death of the minority language (D), and $X = 1$. However, the speakers are still biased towards the vernacular variety. In this case, $\omega^* = 2\alpha-1<0$, and as $\omega = x_1-x_2 < 0$, $x_2>x_1$, meaning that there are more standard variety speakers that prefer the vernacular variety.

Figures~\ref{fig:phasespacefp}c) and~d) are representative cases for coexistence states (C).
We can further study their nature as $s_1$, $s_2$ and $\alpha$ vary. Intuitively, for extreme values of the preference parameter such as $\alpha = 0$ ($\alpha = 1$), coexistence is not possible, as the absence (dominance) of speakers with a preference for the standard variety drives the system to a state with extinction (dominance) of the standard variety. While $\alpha = 0$ or $\alpha = 1$ do certainly not allow for coexistence, $0<\alpha<1$ may allow for it depending on the value of $s_1$ and $s_2$. 

To characterise the phases of the system, we compute the boundaries in phase space which separate the phases of coexistence (C), extinction of the standard variety (E), and its dominance (D). 

To compute the boundaries we simply compare the expressions for the fixed points in Table~\ref{tab:fixedpoints}, as there is one and only one stable fixed point for each parameter configuration. When $X^*_\mathrm{D} = X^*_\mathrm{C}$, the sub-index referring to the ID of the fixed point, we are in the transition line between the dominance of the standard variety and coexistence between the two varieties. In this sense, we obtain the dominance-coexistence (DC) transition line
\begin{equation}
    \label{eq:dctl}
    s_2^\mathrm{DC}(s_1,\alpha) = \frac{s_1-\alpha+s_1 \alpha}{2s_1-\alpha}.
\end{equation}
Similarly for $X^*_\mathrm{E} = X^*_\mathrm{C}$, we obtain the extinction-coexistence (EC) transition line
\begin{equation}
    \label{eq:ectl}
    s_2^\mathrm{EC}(s_1,\alpha) = \frac{(1-\alpha)(s_1-1)}{\alpha + 2(s_1-1)}.
\end{equation}

\begin{figure}[t]
\centering
\includegraphics[width=0.535\textwidth]{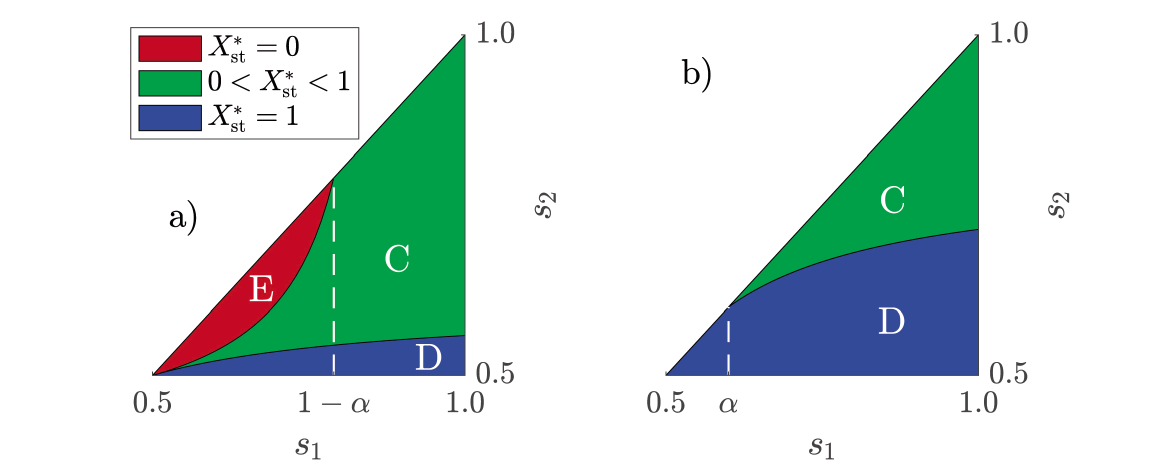}\caption{Phase space for the model with fixed preferences after an analytical analysis, for a) $\alpha < 0.5$ and b) $\alpha > 0.5$, in which the different behaviors of Eqs.~\eqref{eq:dctl} and \eqref{eq:ectl} may be appreciated. The different values of $X$ at the stable fixed point in each parameter configuration, $X^*_\mathrm{st}$, forms 3 phases consisting of dominance of the standard variety (D), its extinction (E), and coexistence of standard and vernacular varieties (C).}
\label{fig:stdiaganalybfafteralpha}
\end{figure}

Eqs.~\eqref{eq:dctl} and~\eqref{eq:ectl} are both null at $s_1 = 0.5$, Eq.~\eqref{eq:dctl} intersects with $s_2(s_1,\alpha) = s_1$, a limit of the phase space, at $s_1 = \alpha$ and Eq.~\eqref{eq:ectl} does so at $s_1 = 1-\alpha$. This means that when $\alpha < 0.5$ only Eq.~\eqref{eq:ectl} intersects with the border $s_2 = s_1$; when $\alpha > 0.5$, only Eq.~\eqref{eq:dctl} exists within the limits of the phase plane and it will intersect with the border $s_2=s_1$. We have then a clear distinction of the phase space depending on whether $\alpha < 0.5$ or $\alpha > 0.5$. This may be seen in Fig.~\ref{fig:stdiaganalybfafteralpha}, where we plot the different values of $X$ in the stable fixed point for each parameter configuration, $X^*_\mathrm{st}$. These values form the phase space for two general values of the preference, $\alpha < 0.5$ and $\alpha > 0.5$. Only for $\alpha < 0.5$ a phase with extinction of the standard variety exists, and in the case of $\alpha > 0.5$, the greater $\alpha$, the smaller the area of coexistence.

\begin{figure*}[t]
\centering
\includegraphics[width=0.7\textwidth]{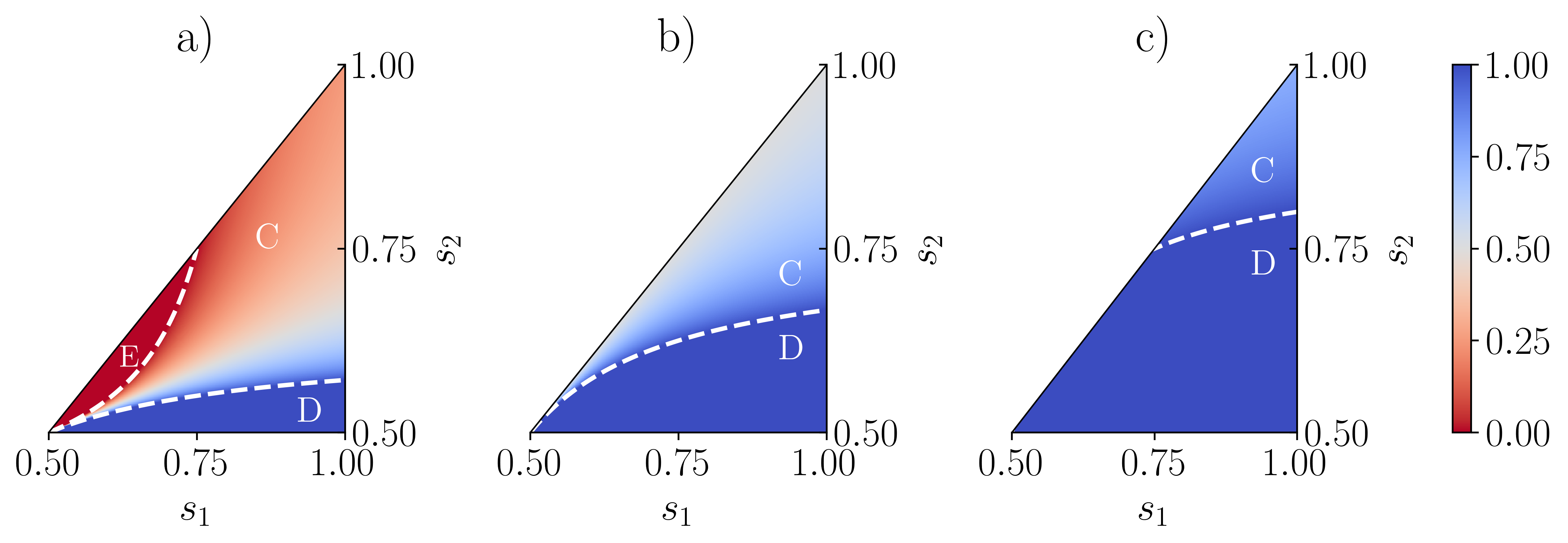}
\caption{Phase space for three different reference values of $\alpha$. We plot the value of $X$ in the stable fixed point following the colour scale in the right for a) $\alpha = 0.25$, b) $\alpha = 0.5$ and c) $\alpha = 0.75$. This value and the fixed point to which it corresponds change depending on the values of $s_1$, $s_2$ and $\alpha$. In dashed lines we plot the transition line from one fixed point to another (see Table~\ref{tab:fixedpoints}), which delimits the region of coexistence of speakers of the two varieties.}
\label{fig:stdiagfpext}
\end{figure*}

To better illustrate the influence of $\alpha$, in Fig.~\ref{fig:stdiagfpext} we plot the boundaries of the phase space and the value of $X$ in the stable fixed point for each parameter configuration for three chosen values of $\alpha$. Fig.~\ref{fig:stdiagfpext}a) depicts a situation with $\alpha = 0.25$, i.e., a quarter of the population prefers the standard variety. This allows for the existence of three regions: the extinction of the standard variety if $s_1$ is sufficiently low and $s_2$ is sufficiently close to $s_1$, the dominance of the standard variety if $s_2$ is sufficiently low, and coexistence of both varieties for a wide range of values of $s_1$ and $s_2$, with a predominant use of the vernacular variety over the standard one.

Interestingly, Figs.~\ref{fig:stdiagfpext}b) and c), which account for $\alpha = 0.5$ and $\alpha = 0.75$, respectively, only show regions in which there exists either coexistence of the two varieties or domination of the standard variety. These figures allow us to observe another direct effect of preferences. As $\alpha \geq 0.5$ in Figs.~\ref{fig:stdiagfpext}b) and c), $0.5\leq X^*_\mathrm{st}<1$ in the coexistence zone: vernacular speakers will, at best, equal in number the standard speakers. Zones of coexistence in which standard speakers outnumber vernacular speakers are no longer allowed, in contrast to Fig.~\ref{fig:stdiagfpext}a). Additionally, the region of coexistence in Fig.~\ref{fig:stdiagfpext}c) has a considerably greater area than the one in Fig.~\ref{fig:stdiagfpext}b), which suggests that a higher preference for the standard variety has a negative effect on the number of possible parameter configurations which allow for coexistence.  

From Eqs.~\eqref{eq:dctl} and \eqref{eq:ectl} we can compute the area of the parameter space with coexistence, $\sigma^c_\mathrm{st}$. A straightforward integration yields
\begin{align}
    \label{eq:coexpropbf05}
    \notag
    &\left( \frac{1}{2}\right)^3 \sigma_\mathrm{st}^\mathrm{c}  = \int_{\frac{1}{2}}^{1-\alpha} s_2^\mathrm{EC}(s_1,\alpha) \mathrm{d}s_1+ \int_{1-\alpha}^1 s_1 \mathrm{d}s_1  \\ 
    & - \int_{\frac{1}{2}}^1 s_2^\mathrm{DC}(s_1,\alpha) \mathrm{d}s_1 = \frac{1}{4}(1-\alpha) \alpha \log \left( \frac{2-\alpha}{\alpha}\right)
\end{align}
for $\alpha < 0.5$, and
\begin{align}
    \label{eq:coexpropaf05}
    \notag
  &\left( \frac{1}{2}\right)^3 \sigma_\mathrm{st}^\mathrm{c} = \int_\alpha^1 s_1 \mathrm{d}s_1   - \int_\alpha^1 s_2^\mathrm{DC}(s_1,\alpha) \mathrm{d}s_1\\ 
  &= \frac{1}{2} (\alpha-1) \alpha \mathrm{atanh}(1-\alpha)
\end{align}
for $\alpha > 0.5$.

\begin{figure}[t]
\centering
\includegraphics[width=0.49\textwidth]{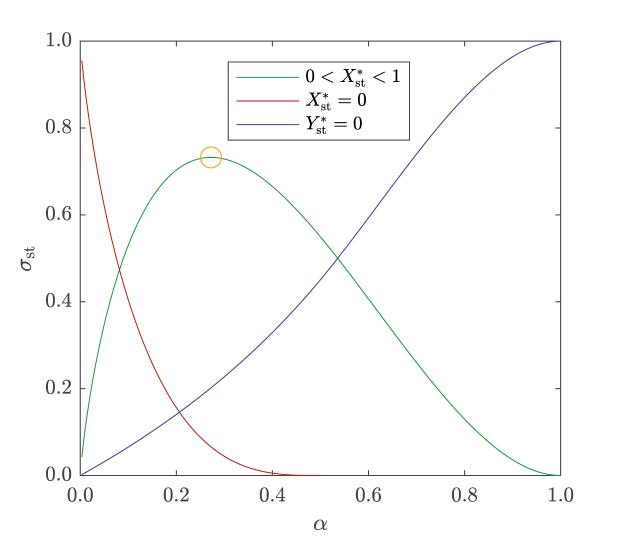}\caption{Area in the parameter space of the stable steady states which imply coexistence, dominance or extinction of one of the two respective varieties, for each value of $\alpha$. The maximum of the coexistence curve is located before $\alpha = 0.5$, as the speakers' preference for the vernacular variety counteracts the different prestige of the varieties. The total area of the parameter space is normalized to 1, as indicated in Eqs.~\eqref{eq:coexpropbf05} and \eqref{eq:coexpropaf05}.}
\label{fig:coexpropanalytical}
\end{figure}

Eqs.~\eqref{eq:coexpropbf05} and \eqref{eq:coexpropaf05} are plotted in Fig.~\ref{fig:coexpropanalytical}.

We would expect that the maximum proportion of situations of stable coexistence occurs for $\alpha = 0.5$, as the proportion of speakers with a preference for one variety or the other would be equal. Nevertheless, we may notice several relevant facts in Fig.~\ref{fig:coexpropanalytical}, where we plot the area in parameter space with $0 < X^*_\mathrm{st} < 1$, implying coexistence; with $Y^*_\mathrm{st} = 0$ implying the dominance of the standard variety, or with $X^*_\mathrm{st} = 0$ implying its extinction, as a function of $\alpha$. 

Firstly, the maximum of the curve for $0 < X^*_\mathrm{st} < 1$ is located at $\alpha < 0.5$. This makes sense, as the standard variety has a higher prestige than the vernacular one. Because of this, coexistence occurs more probably at values of $\alpha$ which imply a higher preference for the vernacular variety than for the standard one, i.e., $\alpha < 0.5$. The preference acts as a counterforce against the differences in prestige, and its effect is maximum at $\alpha_\mathrm{max} = 0.27$.

Secondly, for $\alpha > 0.5$ we stop finding stable fixed points in which $X^*_\mathrm{st} = 0$, as there are more people who prefer the standard variety than the vernacular one, and this in addition to the difference in prestige, prevents the standard variety from losing all of its speakers.

Finally, the curve for the extinction of vernacular variety never vanishes except for $\alpha = 0$. This happens because of the fixed hierarchy of prestiges, i.e., $s_1 > s_2$ holds true at all times. As the standard variety is always more prestigious than the vernacular one, it does not matter how high the preference for the vernacular variety is among the speakers: there will always be a set of parameters which lead to stable situations in which the standard variety dominates.

This is due to the fact that, as the standard prestige is always higher than the vernacular one, as low as the fraction of speakers with a preference for the standard variety is, it is enough to get to stable situations in which $X^*_\mathrm{st} = 1$.

Even though the maximum of the curve is around $\alpha = 0.27$, the specific range of $\alpha$ in which coexistence is stable considering a particular parameter configuration depends on $s_1$ and $s_2$, following Table~\ref{tab:fixedpoints} and Eq.~\eqref{eq:alpharange}. 

To sum up, within a model with a large number of speakers interacting all-to-all, the existence of internal preferences due to the speakers' ideology brings the possibility of coexistence between the two varieties with different prestige. This result agrees with the sociolinguistic situation of many countries and regions where different speech communities show distinct language attitudes. However, societies are not generally made up of completely interconnected speech communities. A more realistic approach takes into account different degrees of coupling between speech communities.

\section{DEPENDENCE ON COUPLING STRENGTH}

Now, we want to address the following question: How does the level of connection between people with different preferences impact how the system behaves?
In other words, we want to investigate the effects of varying degrees of interaction between individuals who have diverse preferences on the overall dynamics of the model.

To do so, we propose a modification of the model with the implementation of a degree of interconnectivity, $\gamma$. This parameter $\gamma$ represents the proportion of all possible links between speakers with different preferences which are actually taking place.

\begin{figure}
\centering
\includegraphics[width=0.5\textwidth]{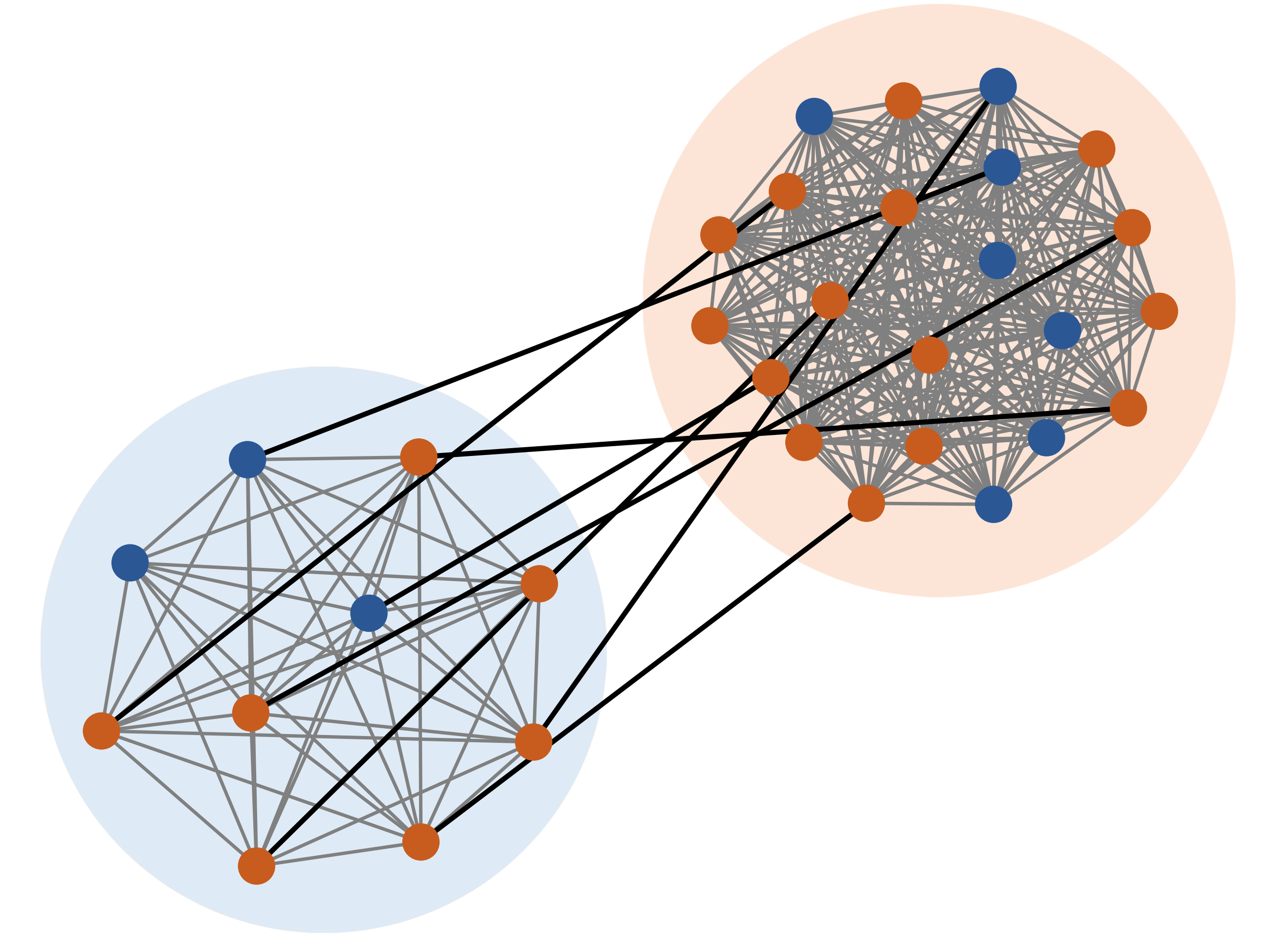}\caption{Illustration of two interconnected networks of varying sizes with coupling between each other. The networks consist of two fully connected speech communities, where individuals share a common preference. Each community is visually depicted by a distinct background color, reflecting the preference for a specific variety. Notably, individuals in these communities have the freedom to use either the standard or vernacular variety for communication, regardless of their preference, as indicated by the node colors. These two communities are partially interconnected by virtue of a fraction of active links $\gamma$, which constitute the lines in black color.}
\label{fig:conexample}
\end{figure}

The situation is depicted in Fig.~\ref{fig:conexample}. The system is made up of two networks. The speakers of each network have exclusively one preference, i.e., we have a community exclusively of speakers who prefer the standard variety and another community exclusively of speakers who prefer the vernacular variety. According to Eq.~\eqref{eq:ligadura1}, the size of the community with a preference for the standard variety in terms of the size of the total population is $\alpha$, and the size of the other community is $1-\alpha$.

To study the dynamics of the system analytically, we assume that each community is fully connected. We can then approximate the dynamics by applying the mean-field approach that we followed in Eqs.~\eqref{eq:s1}-\eqref{eq:s4}, rescaling the interactions between speakers with different preferences by a factor $\gamma$ as an approximation. The new rates then become

\begin{equation}
    \label{eq:tp1c}
    P_{x_1 \rightarrow y_1} = (1-s_1)(y_1+\gamma y_2),
\end{equation}
\begin{equation}
    \label{eq:tp2c}
    P_{x_2 \rightarrow y_2} = s_2(\gamma y_1+y_2),
\end{equation}
\begin{equation}
    \label{eq:tp3c}
    P_{y_1 \rightarrow x_1} = s_1(x_1+\gamma x_2),
\end{equation}
\begin{equation}
    \label{eq:tp4c}
    P_{y_2 \rightarrow x_2} = (1-s_2)(\gamma x_1+x_2).
\end{equation}

As a consequence, the rate equations read

\begin{align}
    \label{eq:x1eqc}
    \notag
    \frac{\mathrm{d}x_1}{\mathrm{d}t} &= (1-s_1)(x_1-\alpha-\gamma y_2)x_1 \\
    &+s_1(\alpha-x_1)\left[x_1-\gamma(\alpha-1+y_2)\right],
\end{align}
\begin{align}
    \notag
    \frac{\mathrm{d}y_2}{\mathrm{d}t} &= (1-s_2)(\alpha-1+y_2-\gamma x_1)y_2 \\
    \label{eq:y2eqc}
    &-s_2(\alpha-1+y_2)\left(y_2+\gamma(\alpha-x_1)\right),
\end{align}
whereas the equations for $x_2$ and $y_1$ can be obtained from Eqs.~\eqref{eq:ligadura1} and~\eqref{eq:ligadura2}. Alternatively, we can work with the rate equations

\begin{align}
    \label{eq:xeqc}
    \notag
    \frac{\mathrm{d}X}{\mathrm{d}t} &= \frac{1}{2} \left(X\left\lbrace(1-2\alpha)+s_2\left[2(\alpha-1)+(1+\gamma)X-\gamma\right] \right. \right. \\ 
    \notag
    &\left. \left. +s_1\left[2\alpha-(1+\gamma)X+\gamma\right]\right\rbrace + \omega\left\lbrace 2\left[(1-s_1)X \right. \right. \right. \\
    \notag
    &\left. \left. \left. +s_1\alpha-s_2(X+\alpha-1)\right]-(s_1-s_2)(2\alpha-1)\gamma \right. \right. \\
    & \left. \left. +(s_1-s_2)(\gamma-1)\omega-1\right\rbrace\right),
\end{align}
\begin{align}
    \label{eq:weqc}
    \notag
    \frac{\mathrm{d}\omega}{\mathrm{d}t} &= \frac{1}{4} \left\lbrace\left[(1-2s_2)(X-\omega)+(2-s_1-s_2)\gamma(X \right. \right. \\ 
    \notag
    &\left. \left. +\omega)\right]\left[X-\omega+2(\alpha-1)\right]- \left[(s_1+s_2)\gamma(X-\omega) \right. \right. \\
    &\left. \left. +(2s_1-1)(X+\omega)\right](X+\omega-2\alpha)\right\rbrace.
\end{align}

We will study $X$ and $\omega$ to obtain a global overview of the dynamics of the system and $x_1 / \alpha$ and $y_2 / (1-\alpha)$ to get insight into what happens inside each community. These two approaches are equivalent, as our system is described only by two independent variables.

In Table~\ref{tab:fixedpointsc} we show the fixed points of Eqs.~\eqref{eq:x1eqc}-\eqref{eq:weqc}. As in the model without coupling, we find three kinds of fixed points: coexistence of the standard and vernacular varieties (C), extinction of the standard variety (E), and its dominance (D). 

For the study of the stability of the fixed points, in Eq.~\eqref{eq:eig1c} we show the eigenvalues for the Jacobian matrix. The linear stability analysis of the fixed points yields an important result: as in the model without coupling, there exists always a unique stable fixed point. Thus, we can study stability diagrams as in the model without coupling.

\begin{table}
\begin{ruledtabular}
\caption{\label{tab:fixedpointsc}Fixed points for the model with coupling. ID stands for the identifier of each fixed point as in Table~\ref{tab:fixedpoints}, and sub-index in $X^*$ and $\omega^*$ stands for the ID of the fixed point. Ther exists a fourth solution for a fixed point, which is not considered further in our analysis as it lays outside our existence range and is never stable.}
\begin{tabular}{ccc}

ID & $X^*$ & $\omega^*$   \\ \hline
E  & 0 & 0          \\
D  & 1 & $2\alpha-1$  \\
C &
  $X^*_\mathrm{C}$ [Eq. (\ref{eq:xfp3c})] & $\omega^*_\mathrm{C}$ [Eq. (\ref{eq:wfp3c})]  \\ 

  \hline \hline
ID & $x_1^*$ & $y_2^*$   \\ \hline
E & 0  & $1-\alpha$ \\
D & $\alpha$ & 0 \\
C &
  $\left(X^*_\mathrm{C} + \omega^*_\mathrm{C} \right)/2$ & $\left[2(1-\alpha)-X^*_\mathrm{C}+\omega^*_\mathrm{C}\right]/2$ \\ 

\end{tabular}
\end{ruledtabular}
\end{table}

To study the effects of coupling in the phase space of the model we may adopt two approaches.

\begin{figure}[t]
\centering
\includegraphics[width=0.48\textwidth]{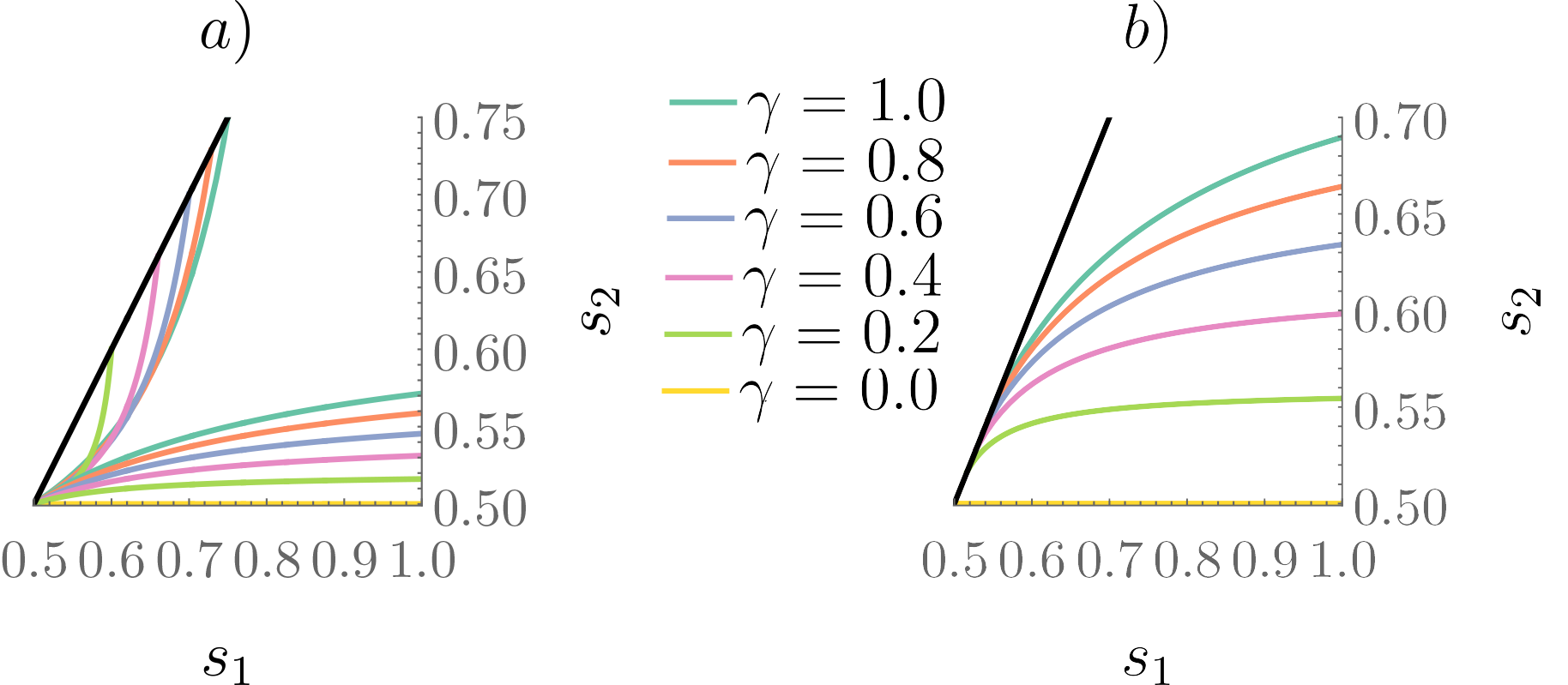}
\caption{Phase diagram of $X$ for a) $\alpha = 0.25 < 0.5$ and b) $\alpha = 0.55 > 0.5$, as in Fig.~\ref{fig:stdiaganalybfafteralpha}, in which we plot the boundaries between the different phases for several values of the coupling parameter $\gamma$. The curves for $\gamma = 1$ reproduce the curves of Fig.~\ref{fig:stdiaganalybfafteralpha}. The black line delimits the phase space allowed by the restrictions over the prestige parameters. Clearly, the increase in the isolation between the two communities enlarges the area in the phase space which allows for coexistence.}
\label{fig:phasediagramsc1}
\end{figure}

Firstly, in Fig.~\ref{fig:phasediagramsc1} we plot the boundaries in $s_1-s_2$ space between the different stable fixed points in terms of $\alpha$ and $\gamma$. We have computed the boundaries by performing numerical solving. Fig.~\ref{fig:phasediagramsc1}a) shows the phase diagram for $\alpha=0.25<0.5$, and Fig.~\ref{fig:phasediagramsc1}b) does so for $\alpha = 0.55>0.5$. Both values of $\alpha$ have been arbitrarily chosen and depict the general behavior of the phase space for $\alpha < 0.5$ and $\alpha > 0.5$, respectively.

The case with $\gamma = 1$, i.e., the case in which both communities are completely connected, is equivalent to the previous model given by Eqs.~\eqref{eq:s1}-\eqref{eq:s4}. Then, we may already make an observation on the influence of the coupling in the dynamics of the system: the decrease of $\gamma$, i.e., the increase in the isolation between the two communities, enlarges the area of the phase space which allows for coexistence. In other words, an increase in the interconnectivity between communities with opposite preferences decreases the area of the parameter space allowing for coexistence. This is an expected result~\cite{Gorenflo2012} that our model captures as a validity check.

Secondly, we compute the boundaries between the different phases characterised by the stable fixed points in the preference-coupling space, i.e., the $\alpha-\gamma$ parameter space, in terms of given values of $s_1$ and $s_2$. The mathematical details of their computation are available in Appendix~\ref{app:ptcoupling}. In Fig.~\ref{fig:analyticalpdseveral} we may show a representative phase space for the model under two different parameter configurations. There are two boundaries which separate coexistence from either vernacular or standard dominance. If we focus on a single value of $\alpha$, the variation of the coupling $\gamma$ allows us to go from one phase to another. For example, let us focus on the case with $s_1 = 0.7$ and $s_2 = 0.6$ of Fig.~\ref{fig:analyticalpdseveral}. Given a fixed value of $\alpha$, we can make a transition from coexistence (C) to standard dominance (D) or from coexistence to vernacular dominance (E). We also have the option to remain in the coexistence phase for every value of $\gamma$. However, for some other values of the prestige parameters, $s_1$ and $s_2$, $\gamma$ allows us to witness more than a single transition. In the case of $s_1 = 0.58$ and $s_2 = 0.51$ of Fig.~\ref{fig:analyticalpdseveral}, for a given set of values of $\alpha$, e.g., $\alpha = 0.15$, we may witness several transitions as we increase $\gamma$: from coexistence to vernacular dominance, then again to coexistence, and then to standard dominance. This is due to the fact that the boundary between standard extinction (E) and coexistence (C) has a local maximum for $\gamma_\mathrm{max}$ at $\alpha_\mathrm{max}$. In Appendix~\ref{app:ptcoupling} we give further details of its calculation and the parameter sets for which this maximum exists.

\begin{figure}[t]
\centering
\includegraphics[width=0.4\textwidth]{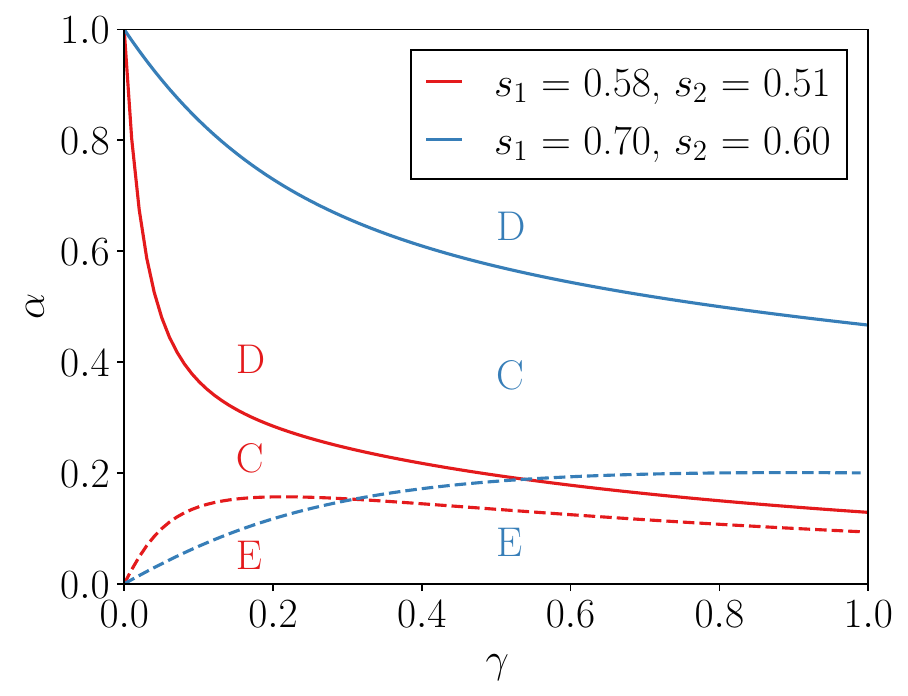}
\caption{Phase diagrams for different values of the prestige parameters $s_1$ and $s_2$. We can see three global phases of coexistence (C), standard variety dominance (D) or vernacular variety dominance, i.e., standard variety extinction (E). Dashed lines represent the boundaries between E and C phases while continuous lines represent the boundaries between C and D phases.}
\label{fig:analyticalpdseveral}
\end{figure}

\subsection{Regime transitions}

As seen in Fig.~\ref{fig:analyticalpdseveral}, some parameter configurations allow us to witness three transitions as the coupling of the two communities increases. In the aforementioned case of $s_1 = 0.58$ and $s_2 = 0.51$, the line $\alpha = 0.15$ crosses the boundaries between phases in three intersection points given by $\gamma_1 = 0.13$, $\gamma_2 = 0.34$ and $\gamma_3 = 0.79$ (see Appendix~\ref{app:ptcoupling} for details about their calculation).

\begin{figure}[t]
\centering
\includegraphics[width=0.49\textwidth]{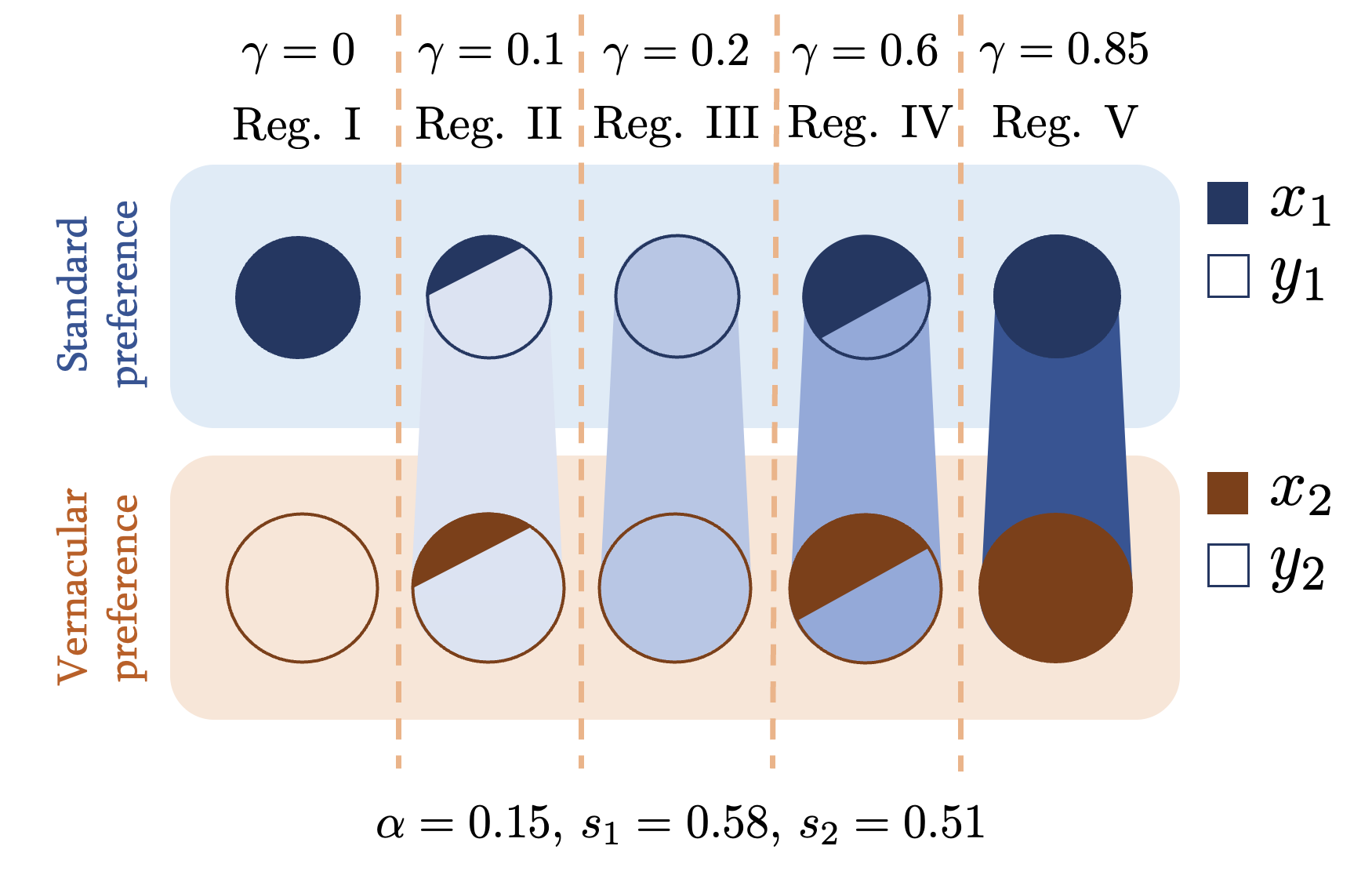}
\caption{Interconnection scheme of the two communities depicted as circles. Empty areas refer to vernacular speakers and filled areas to standard ones. The background colour of the communities differentiates between preferences for either standard or vernacular variety, and the intensity of the colour of the interconnection between the two communities refers to a change in $\gamma$. As the density of interconnections between the two communities increases, the dynamics inside them changes and different regimes are visited. For some values of the preference and prestige parameters, we may witness several transitions through the increase or decrease of $\gamma$. The existence of these transitions is not trivial and leads to re-entering transitions. For example, for $\gamma = 0.1$ coexistence is possible globally and inside each community, but when the coupling is increased to $\gamma = 0.2$, all the speakers switch to the vernacular variety. Again, when we increase the coupling to $\gamma = 0.6$, coexistence is again possible.}
\label{fig:phasedrawing}
\end{figure}

We now discuss in detail all the regimes in which the system may be found when these three intersection points $\gamma_i$ for $i = 1,2,3$ exist. A graphical illustration of these regimes is shown in Fig.~\ref{fig:phasedrawing}.

\begin{itemize}
    \item \textbf{Regime I) null coupling:} In this regime, $\gamma = 0$ and we have two isolated communities in which the only spoken variety is the one preferred by their members. The system is then in phase C, the proportion of standard (vernacular) speakers being determined by the size of their preference community, $\alpha$ ($1-\alpha$). This could describe the situation of an elite that occupies a land but, e.g., does not establish relation with the local people.
    \item \textbf{Regime II) small coupling:} when the coupling is increased to $0 < \gamma \leq \gamma_1$, this little amount of coupling is enough to allow for coexistence due to the influence of each community in the other one. Nevertheless, inside each community, the dominant variety is the majority one. The system remains in phase C. This regime could correspond with the ruling elite increasing the exchanges with the local people. In these cases, there exists a language shift but it is not dramatic.
    \item \textbf{Regime III) medium coupling:} In this regime with $\gamma_1 < \gamma \leq \gamma_2$, the coupling is enough for the majority with less prestige to dominate over the minority with higher prestige. The system is then in phase E. This could correspond to cases such as the Norman elite, who after the England conquest gradually abandoned their more prestigious French language in favor of the English language preferred by the majority.
    Another example would be the rise of Hindi (lower status) versus the decline of English (higher status) in present-day India~\cite{Desilva2020}.
    \item \textbf{Regime IV) reasonably high coupling:} Interestingly, the increase in the coupling for $\gamma_2 < \gamma \leq \gamma_3$ benefits the prestigious minority in comparison to the previous regime. The system re-enters the coexistence phase C and reaches a state in which coexistence is allowed again, but inside each community, the dominant variant is the preferred one among the speakers of the community. There are many examples of this regime nowadays. E.g., in Belgium there are two interacting communities, each keeping their own language.
    \item \textbf{Regime V) strong and total coupling:} with ${\gamma_3 < \gamma \leq 1}$ the coupling is enough for the prestigious minority to dominate in the whole society, so the system arrives to phase D. A historical example of this is the death of many indigenous languages in Latin America, and the survival of Spanish or Portuguese, originally spoken by the ruling minority. Finally, for $\gamma = 1$ we recover the results from the first model (Sec. \ref{sec:firstmodel}).
    
\end{itemize}

\begin{figure}[t]
  \centering
  \begin{minipage}[b]{0.485\textwidth}
    \centering
    \includegraphics[width=\textwidth]{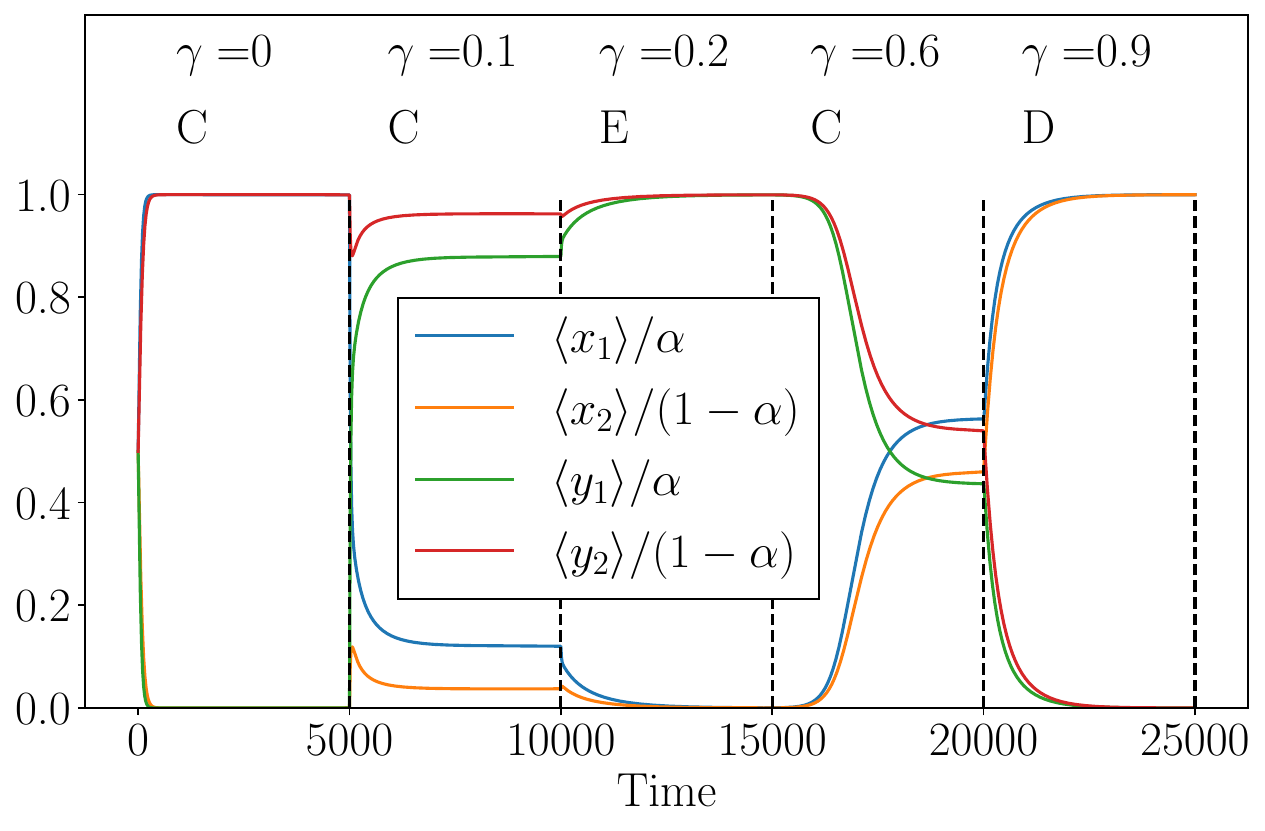}
  \end{minipage}
  \hfill
  \begin{minipage}[b]{0.485\textwidth}
    \centering
    \includegraphics[width=\textwidth]{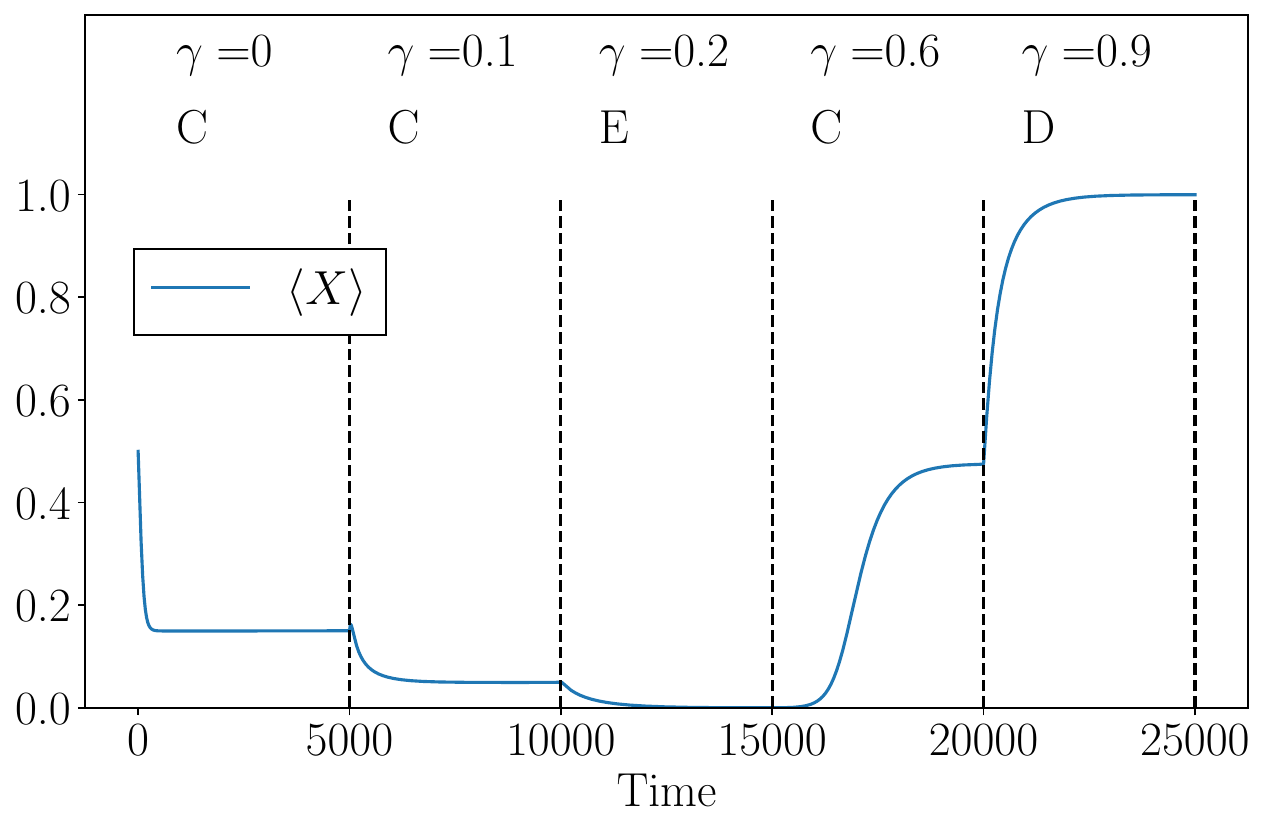}
  \end{minipage}
  \caption{Numerical integration of Eqs.~\eqref{eq:x1eqc} and \eqref{eq:y2eqc} for $s_1 = 0.58$ and $s_2 = 0.51$ showing several transitions. With letters E, C and D we indicate the phases of standard extinction, coexistence of varieties and standard dominance, respectively. By systematically varying the value of $\gamma$, we investigate the transitional behaviour of the system in terms of $x_1$, $x_2$, $y_1$, $y_2$, and $X$. We change the value of $\gamma$ close to the steady state. Since time here is finite, the steady states are never reached and therefore, re-entering the coexistence phase from the extinction one (E-C) is possible. In a), the first four variables are normalised by the size of their corresponding communities. For instance, $x_1/\alpha = 1$ signifies that the entire community, with a size of $\alpha$, predominantly speaks the standard variety. Furthermore, the system's initial state assumes equal distributions of vernacular and standard speakers within each community. }
  \label{fig:numintphases}
\end{figure}

Once the attributes of the different regimes have been described, we will focus on what happens in the system while transitioning from one regime to another using two different approaches. Firstly, by the numerical integration of the rate equations~\eqref{eq:x1eqc} and~\eqref{eq:y2eqc} with abrupt changes of $\gamma$ in time; secondly, by studying the analytical expressions of the stable fixed points in Table~\ref{tab:fixedpointsc} in terms of $\gamma$. Thus, we integrate numerically Eqs.~\eqref{eq:x1eqc} and \eqref{eq:y2eqc} and see how the different regimes are created. One example is shown in Fig.~\ref{fig:numintphases}, where we can see how the increase of the coupling affects both each group of speakers in Fig.~\ref{fig:numintphases}a) and the total proportion of standard speakers in Fig.~\ref{fig:numintphases}b).

The transition from Regime I ($\gamma = 0$ in Fig.~\ref{fig:numintphases}) to Regime II ($\gamma = 0.1$) occurs as a result of a rapid decline in the number of speakers of the standard variety ($x_1$ and $x_2$ in Fig.~\ref{fig:numintphases}a)). This decline is attributed to the small size of the community with a preference for the standard variety and its gradual integration into a much larger community that favors the vernacular variety, as in the example of the Norman conquest. As the interconnection between the two communities increases, the transition from Regime II ($\gamma = 0.1$) to Regime III ($\gamma = 0.2$) leads to vernacular dominance, despite the higher prestige of the standard variety. These regime changes are a direct effect of the interconnection between the two communities. 

However, when the interconnection reaches a sufficiently high level, an interesting re-entering transition from vernacular dominance to a coexistence phase (Regime IV with $\gamma = 0.6$) takes place. This transition is characterised by a rapid decrease in the number of speakers of the vernacular variety and a simultaneous increase in the number of speakers of the standard variety. Surprisingly, the increase in interconnection now has the opposite effect: even though the community with a preference for the vernacular variety is larger than the community favouring the standard variety, the higher prestige of the standard variety noticeably impacts the community with a preference for the vernacular variety, as we can see in the rapid increase in $x_2$ and decrease in $y_2$. 
Finally, the transition from Regime IV ($\gamma = 0.6$) to Regime V ($\gamma = 0.9$) demonstrates a clear dominance of $x_2$ and $x_1$ over $y_2$ and $y_1$, respectively, as in the case of Latin American countries.

\begin{figure*}[t]
  \centering
  \begin{minipage}[b]{0.35\textwidth}
    \centering
    \includegraphics[width=\textwidth]{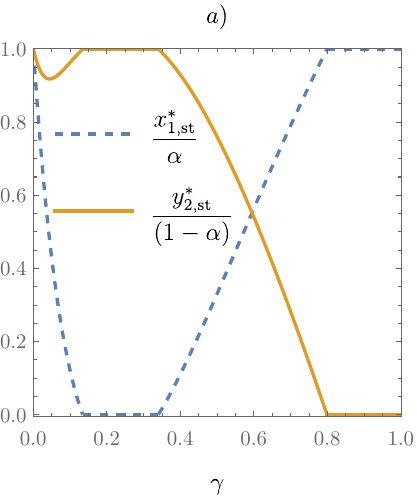}
  \end{minipage}
  \hspace{3em}
  \begin{minipage}[b]{0.35\textwidth}
    \centering
    \includegraphics[width=\textwidth]{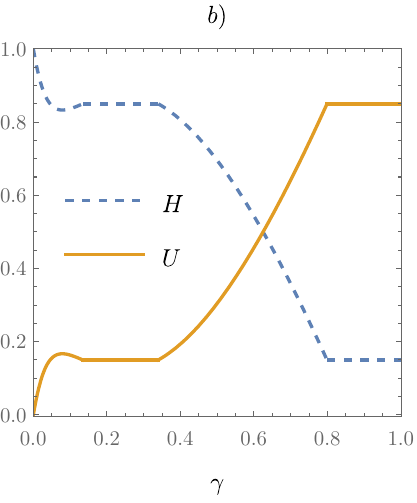}
  \end{minipage}
  \caption{a) Stable fixed points $x^*_{1,\mathrm{st}}$ and $y^*_{2,\mathrm{st}}$ scaled by the sizes of the communities with different preferences in terms of $\gamma$ and b) evolution of the linguistic concordance and disconcordance with the increase of the coupling, for $s_1 = 0.58$, $s_2 = 0.51$ and $\alpha = 0.15$.}
  \label{fig:steadystatesgamma}
\end{figure*}

To understand how the system reaches the aforementioned transitions, we may study the variation of the stable fixed point with $\gamma$ for fixed values of $s_1$, $s_2$ and $\alpha$. In Fig.~\ref{fig:steadystatesgamma}a) we plot the value of the stable fixed point in terms of $\gamma$. As $\gamma$ increases, we observe the aforementioned phases of coexistence, vernacular dominance and standard dominance, through the change from one regime to another. The interest here relies on the evolution of the steady state while reaching each different phase. For that, we also define linguistic concordance $H$ as 
\begin{equation}
    H = x_1 + y_2,
\end{equation}
which clearly refers to the proportion of speakers who are satisfied because their language and preference are aligned (see the faces of the polygons in Fig.~\ref{fig:scheme}). The linguistic disconcordance may be defined as $U = 1-H$. These quantities are plotted in Fig.~\ref{fig:steadystatesgamma}b).

The first observation we can make is that the transitions from one phase to another are smooth. The evolution of linguistic concordance shows that an increase in the coupling causes a decrease in linguistic concordance, as the dynamics of the system rely on the eagerness of the speakers to neglect their preferences. However, there is a narrow region just before Regime III (the phase in which everyone speaks $Y$, so that $x^*_{1,\mathrm{st}} = 0$ and $y^*_{2,\mathrm{st}} = 1-\alpha$) in which linguistic concordance increases. This is due to the fact that the coupling and the relative sizes of the communities allow, in virtue of $P_{x_1 \rightarrow y_1}$ and $P_{x_2 \rightarrow y_2}$ [Eqs.~\eqref{eq:tp1c} and~\eqref{eq:tp2c}, respectively], for a flow from $x_1$ to $y_1$ and then from $x_2$ to $y_2$. As $1-\alpha \gg \alpha$ with $\alpha = 0.15$, the increase of the speakers of $y_2$ has a greater impact in $H$ than the decrease of speakers of $x_1$ and the system gets happier to reach the phase with vernacular dominance.

\begin{figure*}[t]
  \centering
  \begin{minipage}[b]{0.49\textwidth}
    \centering
    \includegraphics[width=\textwidth]{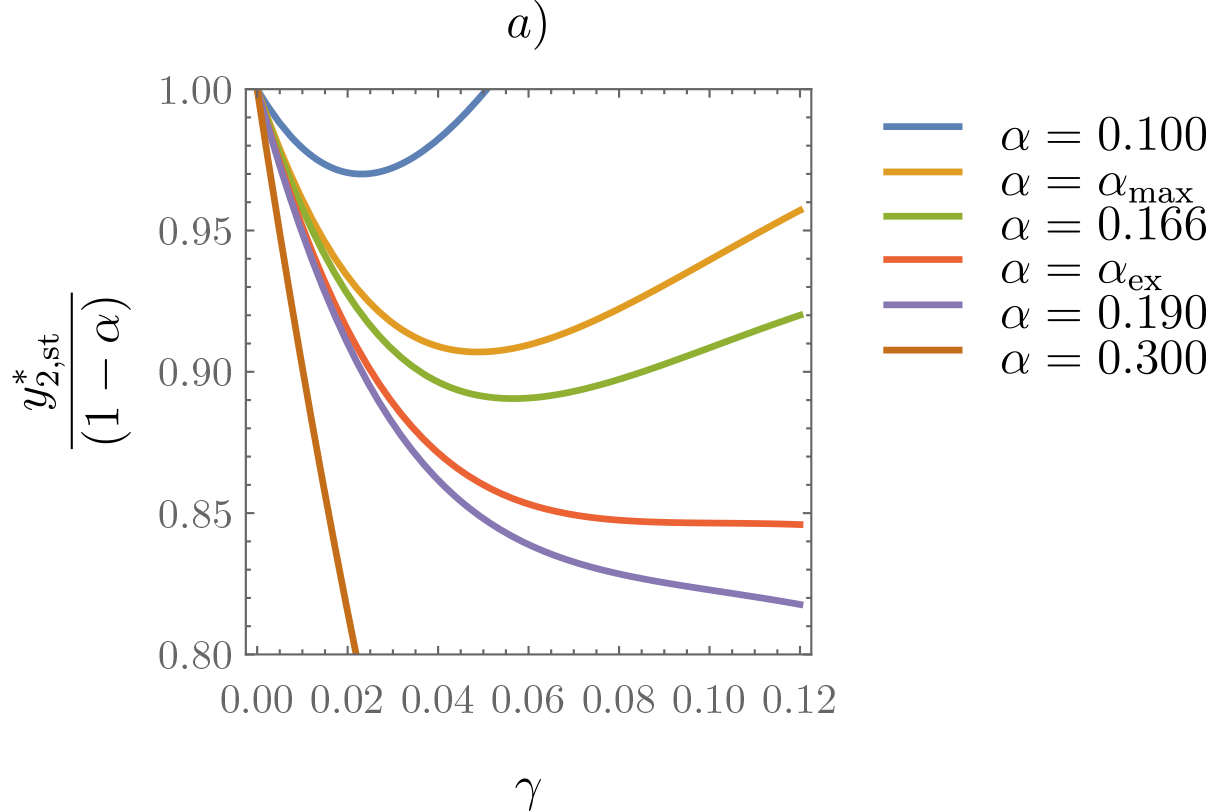}
  \end{minipage}
  \hspace{3em}
  \begin{minipage}[b]{0.35\textwidth}
    \centering
    \includegraphics[width=\textwidth]{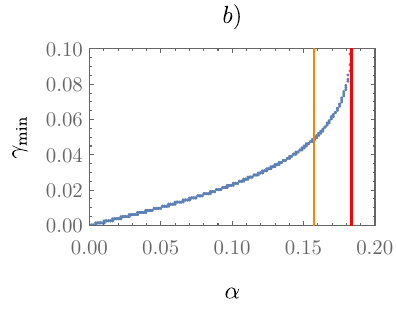}
  \end{minipage}
  \caption{a) Plot of $y^*_{2,\mathrm{st}}$ in terms of $\gamma$ for different values of $\alpha$ and b) location of the minimum --- if existing --- of $y^*_{2,\mathrm{st}}$ in terms of $\alpha$. Both a) and b) are computed for $s_1 = 0.58$ and $s_2 = 0.51$. The existence of a minimum in $y^*_{2,\mathrm{st}}$ is not subjected to the fact that $\alpha < \alpha_\mathrm{max} = 0.157$, as the value of $\alpha$ for which the minimum stops existing, namely $\alpha_\mathrm{ex} = 0.184$, is greater than $\alpha_\mathrm{max}$.}
  \label{fig:minsteadystates}
\end{figure*}

However, reaching a phase with vernacular dominance is not needed for this phenomenon of linguistic concordance momentarily increasing to take place. As we see in Fig.~\ref{fig:minsteadystates}a), even for an $\alpha$ greater than the one corresponding to the maximum for which we can observe the vernacular dominance phase, namely $\alpha_\mathrm{max}$, there exists a minimum in $y^*_{2,\mathrm{st}}$. This is due to the relative size of the communities. The minimum in $y^*_{2,\mathrm{st}}$ is located at $\gamma_\mathrm{min}$, which increases with $\alpha$ as seen in Fig.~\ref{fig:minsteadystates}b). Once the minimum in $y^*_{2,\mathrm{st}}$ fails to exist, linguistic concordance is absolutely decreasing as the coupling increases.

In summary, the analytical exploration of different phases and transitions in the mean-field model with coupling lays the groundwork for understanding the dynamics of societies with languages in contact. However, this approach is limited by the fact that societies have a finite number of speakers. To account for finite-size effects and intricate details, we complement this analysis with an agent-based model implemented on complex networks. This approach allows us to validate the analytical analysis and investigate the influence of network structures, providing a comprehensive understanding of the aforementioned dynamics in realistic social contexts.

\subsection{Finite-size effects}
\label{sec:sim}

We have thus far neglected fluctuation effects since populations are assumed to be large. The results of our deterministic approximation are valid in the thermodynamic limit of infinite systems. To model a more realistic substratum, we now proceed by conducting agent-based simulations of the model, implementing coupling in complex networks. By doing so, we can explore and validate our previous findings while also examining finite-size effects on the dynamics of language competition.

To this purpose, we define a network of $N$ nodes constituted by two fully-connected sub-networks (the so-called communities) with a fixed preference for standard or vernacular variety. Their sizes  are $N_1 = \alpha N$ and $N_2 = (1-\alpha) N$, respectively. 

These sub-networks are connected following a random process of link assignment between speakers with different preferences, as in Fig.~\ref{fig:conexample}. For that, we activate a fraction $\gamma$ of all the possible links between speakers with different preferences; the number of active links, $N_a$, is given by
\begin{equation}
    N_a = \gamma N_1 N_2 = \gamma \alpha (1-\alpha) N^2.
\end{equation}

The simulation of the model with coupling in the aforementioned networks takes place as follows. Each Monte-Carlo step of the simulation consists of a sequential update of all the nodes in the network. The change of the state of an agent during an update depends on a transition probability given by Eqs.~\eqref{eq:tp1c}-\eqref{eq:tp4c} but changing the variables of the proportions of speakers by the local densities of each kind of speaker within the neighbourhood of the agent to update, i.e., those agents who are connected with a direct link to the agent to update.

\begin{figure}[ht]
  \centering
    \includegraphics[width=0.4\textwidth]{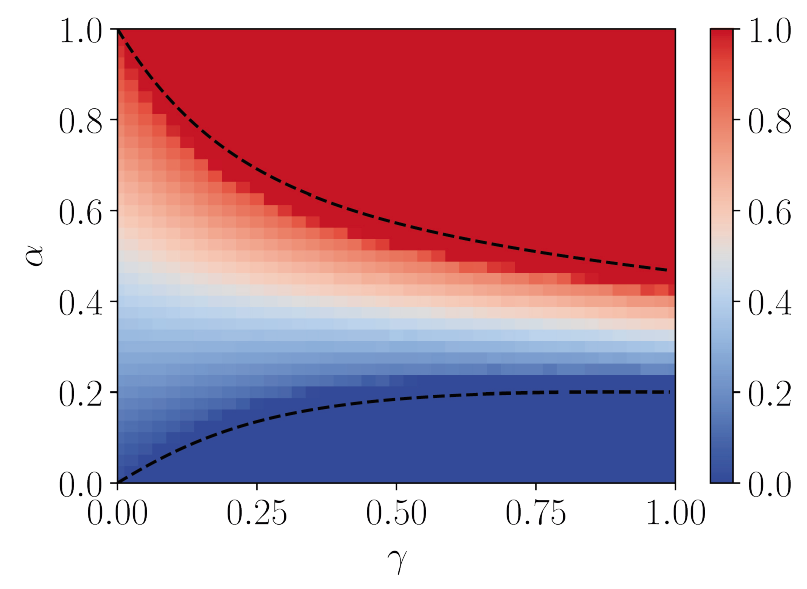}
  \caption{Phase diagram of $X$ for $s_1 = 0.7$, $s_2 = 0.6$. With black, discontinuous lines we plot the boundaries between different phases determined by the value of the stable fixed point for each parameter configuration, as in Fig.~\ref{fig:analyticalpdseveral}. The lines are plotted over the results of a simulation of the model in a complex network of $N = 10^3$ nodes and for $t = 10^4$ Monte-Carlo steps. The color values are calculated as an average over 5 realizations.}
  \label{fig:phasediagramsc}
\end{figure}

In Fig.~\ref{fig:phasediagramsc} we plot the phase diagram for $X$ as a result of simulations with different sets of parameters. We can see a clear agreement between the simulations in networks and the mean-field approach shown in Fig.~\ref{fig:analyticalpdseveral}, meaning that the conclusions drawn from the analytical analysis of the rate equations are valid.

However, a main difference between the rate equations description and the finite size simulation is that the phases D and E are absorbing states of the stochastic dynamics, C is not an absorbing state and a finite size fluctuation will eventually take the system from phase C to either E or D. These absorbing states imply the extinction of either of the varieties, which can have significant societal implications. The relevant question is then: what is the lifetime of phase C for a finite system?

\begin{figure}[t]
\centering
\includegraphics[width=0.45\textwidth]{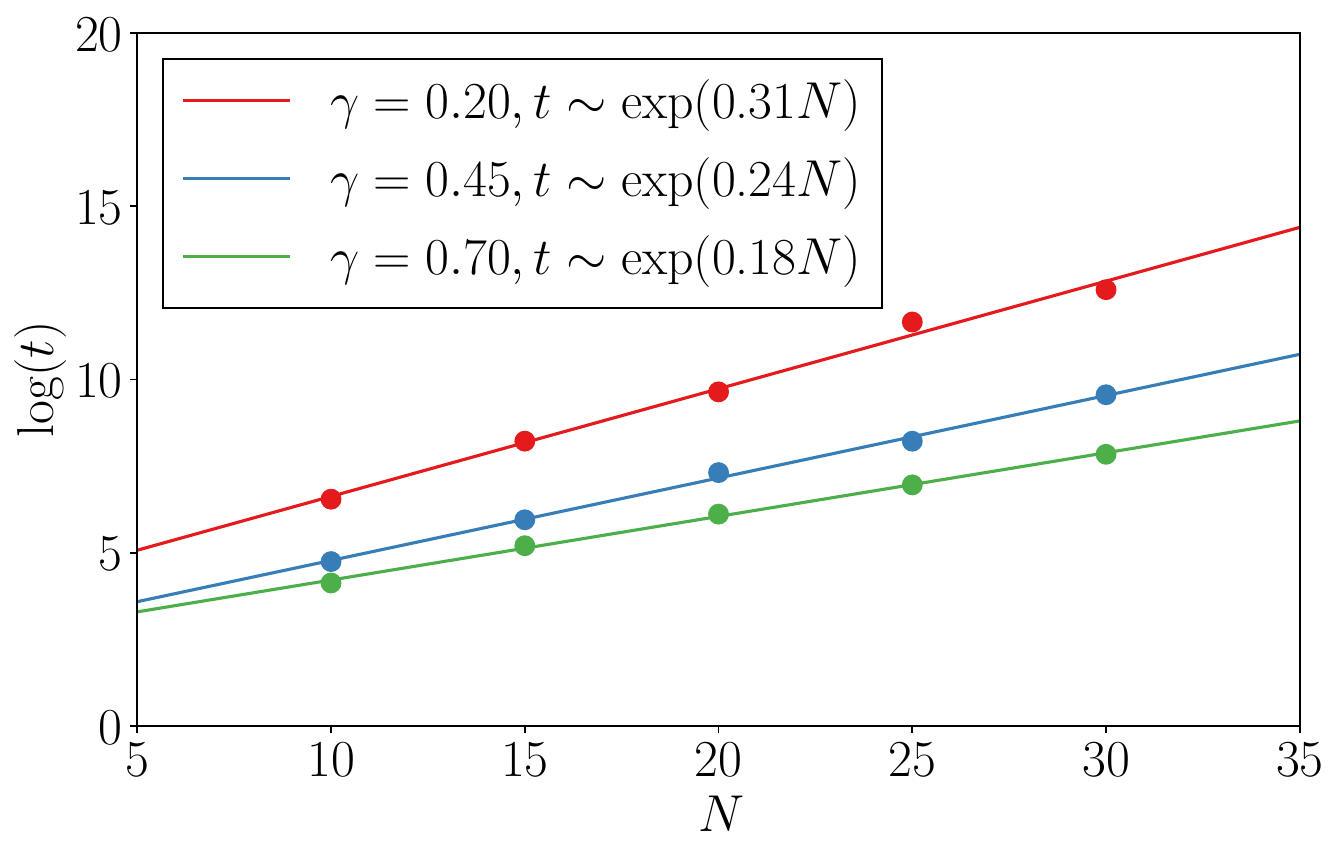}
\caption{Survival time (in Monte-Carlo steps) of coexistence states in terms of network size and coupling strength for $\alpha = 0.4, s_1 = 0.7$ and $s_2 = 0.6$, i.e., the left configuration of Fig.~\ref{fig:phasediagramsc}. We measure the survival time as the first Monte-Carlo step in which the system reaches an absorbing state. Coexistence states survive a time that scales exponentially with network size. The time increases more quickly for higher coupling strengths. The maximum mean relative error is 15 \%.}
\label{fig:survivaltimes}
\end{figure}

As we can see in Fig.~\ref{fig:survivaltimes}, survival times scale exponentially with network size and the exponential growth decreases with coupling, meaning that the coexistence between varieties in a society with a given size has a lifetime which decreases as the interconnection of communities with different preferences or social mixing increases.

\section{Conclusion}

To sum up, we have explored the role of speakers' linguistic preferences in contexts that involve language shift. We did so by proposing a model for two language varieties in contact, accounting for the preferences that speakers may have towards one variety or the other. We have first considered, within a mean field approach, the case of a fully connected population. We have shown that although the standard variety is always more prestigious than the vernacular variety, the speakers' preference quite generally determines the dynamics of the system, allowing for language coexistence in situations in which prestige alone would have led the system towards the extinction of the vernacular variety.

Secondly, we have considered a varying degree of interconnectivity or coupling between the two speech communities with different preferences. The degree of coupling measures the extent to which the two communities communicate with each other. We have found that increasing coupling implies that language coexistence is less likely. This is due to the fact that a stronger connection between speakers with opposing preferences favors the more prestigious variety while reducing the number of individuals aligned with their internal preference.

By increasing the coupling parameter, for fixed prestige values and fixed sizes of the communities with different preferences, we have identified transitions between extinction, dominance, and coexistence phases, which could be applied to real-world scenarios. For example, today's linguistic coexistence in Belgium is allowed in spite of a reasonably high coupling. Additionally, historical sociolinguistic events such as the disappearance of Old French in England or the deaths of many indigenous languages in Latin American countries depend not only on the prestige but also on the coupling degree between the speech communities.

Beyond the mean-field approximation, we have also conducted agent-based simulations of the model on complex networks. These simulations validate the mean-field results and allow for the study of finite-size effects. Remarkably, we have found a nice agreement between the network simulations and the results obtained from the mean-field approximation as far as the behavior with preference and interconnectivity is concerned. We have also found that the lifetime of the coexistence states depends exponentially on system size.

Our model has a number of limitations. First, it considers that the society is spatially homogeneous. However, the varieties spoken in urban and rural areas differ along with their prestige and preferences~\cite{Goncalves2014,Louf2023}. Therefore, there is considerable latitude for the incorporation of the spatial degree of freedom in our model~\cite{Louf2021}. Further, it would be interesting to study the dependence of our results on the interconnectivity within each community. Another limitation is that we do not consider bilingual speakers that are known to alter the transition rates of the model and consequently their fixed points~\cite{Castello2006,Minett2008}. This could be fixed by adding a third population to the dynamics. Finally, we neglect the extent of volatility \cite{Abrams2003, Vazquez2010} and interlinguistic similarity~\cite{Mira2005}, which could be modeled with a parameter scaling the transitions.

More importantly, to achieve predictive power one would require reliable data on language usage evolution and language preference. Available fieldwork data are sparse and restricted to small networks~\cite{Milroy2013}. Social digital datasets have much larger sizes but they are subjected to biases~\cite{Pavalanathan2015,Olteanu2019} and it is not clear to us how to operationalize both language prestige and individual preferences thereof. Nevertheless, this is indeed an interesting research avenue that we plan to explore in the future.

Overall, we highlight the importance of other sociolinguistic parameters beyond the well studied effect of language prestige. In this paper we have discussed the relevant effect of language ideologies and the different degrees of interconnectivity between speech communities. Our findings might have practical implications, especially for policymakers, particularly in the context of minority language preservation and language planning~\cite{Kaplan1997} for contemporary societies.

\appendix
\section{Fixed points and eigenvalues}

The eigenvalues of the first model without coupling (Eqs.~\eqref{eq:s1}-\eqref{eq:s4}) are given by

\begin{align}
    \label{eq:eig1}
    \lambda_{1,2}  &= \frac{1}{2} \left[\alpha (1-s_1)\mp\frac{1}{2} \sqrt{A}+ 3X^*(s_2-s_1)
     \right. \nonumber \\
     & \left. +s_1(1-\omega^*)+s_2(\alpha-\omega^*-2)+\omega^* \right], 
\end{align}
where
\begin{align}
\label{eq:eig2}
    A &= 4 \left(\alpha +s_1 (3X^*+\omega^* -\alpha-1)+s_2 (\omega^*-\alpha -3X+2) \right. \nonumber \\
    &  \left. -\omega^* \right)^2+8 \left(-2 \alpha +s_1 \left(2 \alpha+4 s_2 (1-2 X)^2 \right. \right. \nonumber\\
    &\left. \left. +X^*(7-2 \alpha -8 X^*)-\omega -2\right) -s_2 (-2 \alpha +X^* (2 \alpha \right. \nonumber\\
    &\left. +8 X^*-11)+\omega+4)+X^*(2 \alpha +4 X^*-5)+\omega^* +2\right).
\end{align}

Note that Eqs.~\eqref{eq:eig1} and~\eqref{eq:eig2} depend only on the parameters and on the specific values of $X^*$ and $\omega^*$,
because $Y$ and $z$ can be eliminated using Eqs.~\eqref{eq:ligadura1} and~\eqref{eq:ligadura2}.

The fixed points in Table~\ref{tab:fixedpointsc} are given by

\begin{widetext}
\begin{align}
    \label{eq:xfp3c}
    \notag
    X^*_\mathrm{C}(s_1,s_2,\alpha,\gamma) &=\frac{1}{4 (1-2 s_1)^2 (1-2 s_2)^2 (1+\gamma)} \left[(-1+2 s_1)(1-2 s_2)^2 (-2+2 \alpha-\gamma) \gamma - (4\alpha-2)\sqrt{G(s_1,s_2,\gamma)}\right.\\
    & \left.  + (1-2 s_1)^2 (2 s_2 - 1) (-2+4 s_2 (1+\gamma) + \gamma (-2+2 \alpha+\gamma))\right],
\end{align}
where
\begin{align}
    \notag
    G(s_1,s_2,\gamma) &=(1-2 s_1)^2 (1-2 s_2)^2 ((1-2 s_1)^2 (1-2 s_2)^2 - 2 (-1+2 s_1)(2 s_2 - 1) (-s_2+s_1 (2 s_2 - 1)) \gamma^2 \\
    \notag
    & +(s_1-s_2)^2 \gamma^4),
\end{align}

\begin{align}
    \label{eq:xfp4c}
    \notag
    X^*_4(s_1,s_2,\alpha,\gamma) &=\frac{1}{2 (1-2 s_1)^2 (1-2 s_2)^2 (1+\gamma)} \left[ (1- 2 \alpha )\sqrt{G(s_1,s_2,\gamma)}  +2 s_1^2 (2 s_2 - 1) (-2+4 s_2 (1+\gamma) \right. \\
    \notag
    & \left. + \gamma (-2+2 \alpha+\gamma)) +(2 s_2 - 1)(-1+2 (\alpha-1) \gamma+s_2 (2+\gamma (4-2 \alpha+\gamma))) \right. \\ 
    & \left. - s_1(2 s_2 - 1)(\gamma-4 (6(-1+ \alpha)+\gamma)+2 s_2 (4+\gamma (6-2 \alpha+\gamma)))\right].
\end{align}

\begin{align}
    \label{eq:wfp3c}
    \notag
    \omega^*_{3,4}(s_1,s_2,\alpha,\gamma) &=\frac{\gamma (-2 \alpha (\gamma-1) + \gamma)}{(2s_2-1)} \mp\frac{2\sqrt{H(s_1,s_2,\gamma)}}{(1-2s_1)^2(1-2s_2)^2} \frac{1}{4(\gamma-1)}  \\
    & + \frac{-2 + 4s_1(2\alpha-1)(\gamma-1)+2\alpha(\gamma-2)(\gamma-1) - (\gamma-4)\gamma}{(2s_1-1)},
\end{align}
where
\begin{align}
    \notag
    H(s_1,s_2,\gamma) = \frac{-2 + 4s_1(2\alpha-1)(\gamma-1)+2\alpha(\gamma-2)(\gamma-1) - (\gamma-4)\gamma}{(2s_1-1)}.
\end{align}
\end{widetext}

The eigenvalues of the model with coupling (Eqs.~\eqref{eq:xeqc} and \eqref{eq:weqc}) are given by

\begin{align}
    \label{eq:eig1c}
    \notag
    \lambda_{1,2} &= \frac{1}{2} \left( (s_1+s_2)\left[2(\alpha-\omega^*)+\gamma(\omega^*-\alpha)\right] +s_1 \gamma \right.\\
    \notag
    &\left. - 2 s_2 + \gamma (\alpha-\omega^*-1)+2(\omega^*-\alpha)-(\gamma \right. \\
    &\left. +2)X^*(s_1-s_2)+1\mp \frac{1}{2}\sqrt{A_c} \right),
\end{align}

where
\begin{widetext}
\begin{align}
\notag
    A_c &= 8 \left(-2 (2 s_1 - 1) (2 s_2 - 1) (\alpha-1) \alpha2 \left(-2 s_2 + (\alpha-1)^2 - s_1 (\alpha-1)^2 + s_2 (4 - 3 \alpha) \alpha   \right. \right. \\
    \notag
    & \left. \left. + s_1 s_2 (2  \gamma+ 4 (\alpha-1) \alpha)\right)- 2 (s_1 - s_2) (\alpha-1) \alpha \gamma^2 + 2 (2 s_1 -1) (2 s_2 - 1) (X^*)^2 (1 + \gamma)  \right. \\
    \notag
    & \left. +X^* \left(2 \alpha \gamma-2 - 3 \gamma + s_2 (4 + \gamma (6 - 2 \alpha + \gamma)) - s_1 ( 8 s_2 -4 + (1 + \gamma)+\gamma (2 \alpha -4+ \gamma))\right) \right. \\
    \notag
    & \left. + \left(2 (2 s_1 - 1) (2 s_2 - 1) (2 \alpha -1)+(4 s_1-3) (2 s_2 - 1) \gamma + 2 (s_1 (3 - 8 s_2) + 5 s_2-2) \alpha \gamma  \right. \right.\\
    \notag
    & \left. \left. + (s_1 - s_2) (2 \alpha-1) \gamma^2\right) \omega^* + 2 (2 s_1 - 1) (2 s_2-1) (\gamma-1) (\omega^*)^2\right)+ 4 \left(1 - 2 \alpha - \gamma + \gamma (\alpha - \omega^*) \right. \\
    \notag
    & \left.+ 2 \omega^* + s_1 \left(2 \alpha + \gamma - \alpha \gamma - X^* (2 + \gamma)+ (\gamma-2) \omega^*\right) + s_2 (-2 + 2 \alpha - \alpha \gamma + X^* (2 + \gamma) + (\gamma-2) \omega^*)\right)^2.
\end{align}
\end{widetext}

They only depend on the parameters and on the specific values of the fixed points $X^*$ and $\omega^*$.

\section{Phase transitions due to coupling}
\label{app:ptcoupling}

By computing the analytical expressions for the curves which define the boundaries of the several phases in the $\alpha-\gamma$ parameter space of the model with coupling we can analyze some interesting results.

For that, we revisit the fixed points in Table~\ref{tab:fixedpointsc}. If we focus on the value of $X$, and as there exists one and only one stable fixed point for each parameter configuration, we can compute the boundary of the phases of coexistence (C) and dominance (D) of the standard variety by solving $X^*_\mathrm{D}(s_1,s_2,\alpha,\gamma)=X^*_\mathrm{C}(s_1,s_2,\alpha,\gamma)$, which yields

\begin{widetext}
\begin{align}
    \label{eq:dctlc}
    \notag
    \alpha^\mathrm{DC}(s_1,s_2,\gamma) &=\frac{1 + \sqrt{B(s_1,s_2,\gamma)} + 2s_1^2(2s_2-1)(4s_2(1+\gamma)-2-\gamma(2+\gamma)) + s_1(2s_2-1)(4+\gamma(2+\gamma)+2s_2(\gamma-2)\gamma-4) }{{2((2s_1-1)(s_1+s_2-1)(2s_2-1)\gamma + \sqrt{B(s_1,s_2,\gamma)})}} \\
    &+\frac{s_2(\gamma-4^2-2s_2(\gamma-2^2))}{2((2s_1-1)(s_1+s_2-1)(2s_2-1)\gamma + \sqrt{B(s_1,s_2,\gamma)})},
\end{align}
where
\begin{align}
    \notag
    B(s_1,s_2,\gamma) &=(1-2s_1)^2 (1-2s_2)^2 \left((1-2s_1)^2 (1-2s_2)^2 \right. \\
    \notag
    & \left. -2(2s_1-1)(2s_2-1)(-s_2+s_1(2s_2-1))\gamma^2 + (s_1-s_2)^2\gamma^4\right).
\end{align}
\end{widetext}

We can proceed in the same way for computing the boundary of the phases of coexistence (C) and extinction (E) of the standard variety by solving $X^*_\mathrm{E}(s_1,s_2,\alpha,\gamma)=X^*_\mathrm{C}(s_1,s_2,\alpha,\gamma)$, which yields

\begin{widetext}
\begin{align}
    \label{eq:ectlc}
    \notag
    \alpha^\mathrm{EC}(s_1,s_2,\gamma) &=\frac{2(2s_1-1)s_2^2(2+4s_1(\gamma-1)+(\gamma-4)\gamma)}{{2(2s_1-1)(2s_2-1)(1+s_1(4s_2-\gamma-2)+s_2(\gamma-2))(\gamma-1)}}  \\
    \notag
    &+ \frac{\sqrt{C(s_1,s_2,\gamma)} + (2s_1-1)(1-2\gamma+s_1(\gamma-2(2+\gamma)))}{2(2s_1-1)(2s_2-1)(1+s_1(4s_2-\gamma-2)+s_2(\gamma-2))(\gamma-1)} \\
    &-\frac{ (2s_1-1)s_2(4+(\gamma-8)\gamma+2s1(\gamma-4(4+\gamma)))}{2(2s_1-1)(2s_2-1)(1+s_1(4s_2-\gamma-2)+s_2(\gamma-2))(\gamma-1)},
\end{align}
where
\begin{align}
    \notag
    C(s_1,s_2,\gamma) &=(1-2s_1)^2 (1-2s_2)^2 ((1-2s_1)^2 (1-2s_2)^2 - 2(2s_1-1)(2s_2-1)(s_1(2s_2-1-s_2))\gamma^2\\
    \notag
    & + (s_1-s_2)^2\gamma^4).
\end{align}
\end{widetext}

Firstly, after a long but simple derivation, we can see that 

\begin{equation}
    \notag
    \frac{\mathrm{d}\alpha^\mathrm{DC}(s_1,s_2,\gamma)}{\mathrm{d}\gamma} < 0 \ \forall \gamma,
\end{equation}
hence $\alpha^\mathrm{DC}(s_1,s_2,\gamma)$ is monotonically decreasing. 

Interestingly, $\alpha^\mathrm{EC}(s_1,s_2,\gamma)$ may have a maximum, as 

\begin{equation}
    \notag
    \exists \mathcal{C}\{s_1,s_2\} / \frac{\mathrm{d}\alpha^\mathrm{EC}(s_1,s_2,\gamma)}{\mathrm{d}\gamma} = 0,
\end{equation}
for
\begin{equation}
    \notag
    \gamma_\mathrm{max} = \sqrt{\frac{1-2(s_1+s_2)+4s_1s_2}{s_1-s_2}},
\end{equation}

so that
\begin{equation}
    \alpha_\mathrm{max}(s_1,s_2) = \alpha^\mathrm{EC}(s_1,s_2,\gamma_\mathrm{max}).
\end{equation}

\begin{figure}[t]
\centering
\includegraphics[width=0.3\textwidth]{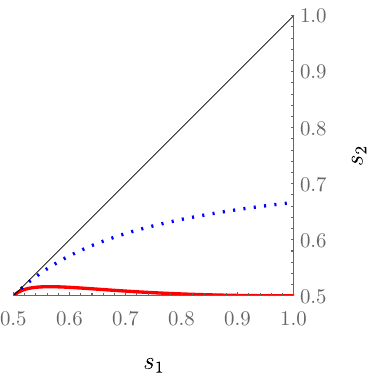}
\caption{The values of $s_1$ and $s_2$ under the dotted curve imply that the standard boundary in the phase space is monotonically decreasing while the vernacular boundary has a maximum. This allows us to find several transitions, as we can go from coexistence (C) to standard extinction (E), then to coexistence (C) and then to standard dominance (D). For the values over the boundary, the phase space consists in a standard boundary which monotonically decreases and a vernacular boundary which monotonically increases. Notably, the parameter configurations allowing for three transitions account for 45\% of the total parameter space. Additionally, the configurations of $s_1$ and $s_2$ under the solid line imply $\Delta \alpha(s_1,s_2) > 0$, meaning that we can witness 3 transitions. These particular parameter configurations cover 3\% of parameter space.}
\label{fig:s1s2pt}
\end{figure}

To compute this set $\mathcal{C}$, we impose $0\leq\gamma_\mathrm{max}\leq1$ and we arrive at the condition 
\begin{equation}
    \notag
    s_2 \leq \frac{3s_1-1}{4s_1-1}.
\end{equation}
This condition defines the region described by the dotted curve in Fig.~\ref{fig:s1s2pt}.

A maximum in $\alpha^\mathrm{EC}(s_1,s_2,\gamma)$ implies the existence of at least 3 different phases for some values of $\alpha$. If the minimum value of $\alpha^\mathrm{DC}(s_1,s_2,\gamma)$, i.e., $\alpha^\mathrm{DC}(s_1,s_2,1)$, is such that $\alpha^\mathrm{DC}(s_1,s_2,1)<\alpha_\mathrm{max}(s_1,s_2)$, we can find 4 phases for 

\begin{equation}
    \alpha \in \big[\alpha^\mathrm{DC}(s_1,s_2,1),\alpha_\mathrm{max}(s_1,s_2)\big].
\end{equation}

If we define the following quantity

\begin{equation}
    \Delta\alpha(s_1,s_2) = \alpha_\mathrm{max}(s_1,s_2)-\alpha^\mathrm{DC}(s_1,s_2,1),
\end{equation}
we have that $\Delta \alpha(s_1,s_2) > 0$ only for certain values of $s_1,s_2$, which are given by

\begin{align}
    \notag
    s2 < &\frac{7 s_1 - 7 s_1^2 + 11 s_1^3-3}{3 (2 s_1 - s_1^2 + 4 s_1^3-1)} + \frac{E(s_1)}{3 (2 s_1 - s_1^2 + 4 s_1^3-1)}  \\
    & +  \frac{4 s_1^{-1} - 8 + 5 s_1 + 2 s_1^2 - 23 s_1^3}{3 (2 s_1- s_1^2 + 4 s_1^3-1)D(s_1)},
\end{align}
where
\begin{align}
    \notag
    D(s_1) &= s_1^3 \left(24 s_1 - 54 s_1^2 + 167 s_1^3 - 261 s_1^4 + 249 s_1^5  \right.\\
    \notag
    & \left. - 181 s_1^6 -8+6 \sqrt{3} s_1 (E(s_1))^\frac{1}{6}\right),
\end{align}
and
\begin{align}
    \notag
    E(s_1) &= 4 - 36 s_1 + 142 s_1^2 - 361 s_1^3 + 726 s_1^4 - 1178 s_1^5 \\
    \notag
    &+ 1518 s_1^6 - 1633 s_1^7 + 1378 s_1^8 - 864 s_1^9 + 416 s_1^{10}.
\end{align}

These values of $s_1$ and $s_2$ are depicted by the solid red line in Fig.~\ref{fig:s1s2pt}.

We are now in a position to compute analytically the values of $\gamma$ in which a given value of $\alpha$, i.e., a horizontal line in $\alpha-\gamma$ phase space, intersects with the vernacular and standard boundaries. For the standard boundary, we have that
\begin{widetext}
\begin{align}
    \notag
    \gamma^s_1 &= \frac{s_1 + (s_2-1)\alpha^2 + F(s_1,s_2,\alpha)}{2(s_1-s_2)(\alpha-1)\alpha}  \\
    &+\frac{\sqrt{4(2s_1-1)(s_1-s_2)(2s_2-1)(\alpha-1)^2\alpha^2 + \left[s_1 + (s_2-1)\alpha^2 + G(s_1,s_2,\alpha)\right]^2}}{2(s_1-s_2)(\alpha-1)\alpha},
\end{align}
\end{widetext}
where
\begin{align}
    F(s_1,s_2,\alpha) &= s_1\alpha(3\alpha-2) - s_1 s_2\left[2 + 4(\alpha-1)\alpha\right].
\end{align}

For the vernacular boundary, we have performed numerical solving.

\begin{acknowledgments}
This work was partially supported by the Spanish State Research Agency
(MCIN/AEI/10.13039/501100011033) and FEDER (UE) under project APASOS (PID2021-122256NB-C21) and the Mar{\'\i}a de Maeztu project CEX2021-001164-M, and by the Government of the Balearic Islands CAIB fund ITS2017-006 under project CAFECONMIEL (PDR2020/51).
\end{acknowledgments}

\bibliography{biblio}

\end{document}